\documentclass[english,aps,prc,floats,floatfix,nofootinbib]{revtex4}

\usepackage[T1]{fontenc}
\usepackage[latin9]{inputenc}
\usepackage{amsmath}
\usepackage{graphicx}
\usepackage{amssymb}

\makeatletter
\@ifundefined{textcolor}{}
{
 \definecolor{BLACK}{gray}{0}
 \definecolor{WHITE}{gray}{1}
 \definecolor{RED}{rgb}{1,0,0}
 \definecolor{GREEN}{rgb}{0,1,0}
 \definecolor{BLUE}{rgb}{0,0,1}
 \definecolor{CYAN}{cmyk}{1,0,0,0}
 \definecolor{MAGENTA}{cmyk}{0,1,0,0}
 \definecolor{YELLOW}{cmyk}{0,0,1,0}
}

\usepackage{bm}
\makeatletter
\usepackage{graphics}
\usepackage{epsf}
\makeatother

\def\s#1{\setbox0=\hbox{$#1$}  
   \dimen0=\wd0     
   \setbox1=\hbox{/} \dimen1=\wd1  
   \ifdim\dimen0>\dimen1   
      \rlap{\hbox to \dimen0{\hfil/\hfil}} 
      #1     
   \else     
      \rlap{\hbox to \dimen1{\hfil$#1$\hfil}} 
      /      
   \fi}      %

\newcommand{\nn}{\nonumber}

\makeatother
\usepackage{babel}

\begin{document}

\title{Background field approach to electromagnetic properties of baryons}

\author{Tim Ledwig}
\email{ledwig@kph.uni-mainz.de}
\affiliation{Institut f\"ur Kernphysik, Universit\"at Mainz, D-55099 Mainz, Germany}

\begin{abstract}
  We investigate the self-energies of particles in an external magnetic
  field $B$. The dependence is generally of the type $\sqrt{P\left(B\right)}$
  with $P$ a polynomial in $B$ and the participating masses. The non-analytic
  point depends on the mass and charge constellation, is unproblematic
  for stable particles but constrains the linear energy shift approximation
  for resonances. We recover an earlier reported condition 
  on when the energy can be expanded in $B$ and derive two more conditions.
  Further, we obtain the $B$ dependent self-energies of the
  nucleon and $\Delta(1232)$-isobar in the $SU(2)$ covariant chiral
  perturbation theory.
\end{abstract}


\maketitle

\section{Introduction}

Electromagnetic (EM) properties of particles are fundamental observables
for their internal structure. For hadrons it is still not possible
to reveal this structure analytically from QCD first principles. Two
prominent ways to obtain information on baryon EM properties are chiral
perturbation theories ($\chi$PT) and lattice QCD (lQCD) which both
allow to investigate the quark/pion mass dependence of observables.
At present, finite volume lQCD results are mainly compared to infinite
volume $\chi$PT ones 
\cite{Alesandrou:NlQCD,Bratt:NlQCD,Collins:NlQCD,Syrisyn:NlCQD,Ledwig(2011):covChptNDelta}
where discrepancies in the small pion mass region are seen 
\cite{Syrisyn:NlCQD,Collins:NlQCD,Ledwig(2011):covChptNDelta}.

EM finite volume effects on the $\chi$PT side are therefore of interest, however, 
reveal certain subtleties. Special care has to be taken for the decomposition
of the vector-current matrix element in form factors as well as for their Lorentz
invariance \cite{Tiburzi3p}. An alternative approach, e.g. to investigate the
magnetic moment, is to use the particle's self-energy in an external EM
field. This was done in \cite{TiburziBFT} for pions and nucleons in the non-relativistic
heavy baryon $\chi$PT. 

One motivation for this work is to derive the corresponding covariant infinite
volume $\chi$PT self-energies of the nucleon and $\Delta(1232)$ that would be needed
for a future finite volume study. Another one is to investigate further the non-analytic EM field dependence of
self-energies \cite{Ledwig(2010):nonAna}. We see that beside the found condition
for expanding a resonance self-energy in the field strength there exist two more.
For this, we modernize the EM background field technique (BFT) of \cite{Sommerfield}
and apply it to stable and unstable particles.

In 1958 C. M. Sommerfield used the BFT to obtain the electron's AMM
up to the fourth order, $\kappa_{e}=\frac{e^{2}}{8\pi^{2}}-0.328\frac{e^{4}}{4\pi^{3}}$,
correctly for the first time. However, it is the well known three
point function method, i.e. the one-photon approximation, that became
the preferred method to calculate EM moments in field theories. We
will use both techniques to check our formulas. It turns out that the
self-energies of particles, stable as well as unstable, generally
depend non-analytically on the magnetic field $B$. This is e.g.
seen for the nucleon and $\Delta\left(1232\right)$-isobar in the
covariant $SU(2)$ B$\chi$PT with $N-\pi$ loops where the $\tilde{MS}$
renormalized results with $\mu=m_{\pi}/M_{N}$ are:

\begin{eqnarray}
  \Sigma_{p}\left(\mathcal{B}\right) & = & \frac{M_{N}C_{N}}{24\left(1
    -\mathcal{B}\right)^{4}}\left[n_{1}+n_{2}\ln\mu+\frac{n_{3}}{\sqrt{4\mu^{2}
        -\left(\mathcal{B}+\mu^{2}\right)^{2}}}\text{\ensuremath{\arccos}}
    \frac{\mathcal{B}+\mu^{2}}{2\mu}\right]\,\,\,,\\
  \Sigma_{\Delta^{+}}\left(\mathcal{B}\right) & = & 
  \frac{M_{\Delta}C_{\Delta}}{\left(1-\mathcal{B}\right)^{4}}\Big[d_{1}+d_{2}
    \ln\mu+d_{3}\ln r+d_{4}\ln\left(1-\mathcal{B}\right)\nn\\
    &  & +\frac{\left(\mathcal{B}-1\right) d_{5}\mbox{arctanh
        \ensuremath{\omega}}_{1}}{\sqrt{\lambda^{2}+4\mu^{2}\mathcal{B}}}
    +\frac{\left(\mathcal{B}-1\right)d_{6}\mbox{arctanh
        \ensuremath{\omega}}_{2}}{\sqrt{\lambda^{2}+\mathcal{B}\left(\mathcal{B}
        -2+2r^{2}+2\mu^{2}\right)}}+d_{7}\ln\left(r^{2}-\mathcal{B}\right)\Big],
\end{eqnarray}
with certain polynomials $n_{i}=n_{i}\left(\mathcal{B},\mu\right)$,
$d_{i}=d_{i}\left(\mathcal{B},\mu,r\right)$, $\lambda^{2}=\lambda^{2}\left(\mu,r\right)$
and further non-analytic functions $\omega_{i}=\omega_{i}\left(\mathcal{B},\mu,r\right)$
and notations given later. One difference of the self-energies is
that the square root in $\Sigma_{p}\left(B\right)$ can be expanded
in a weak magnetic field $B$ for all pion masses $m_{\pi}>0$ whereas
for the $\Delta\left(1232\right)$-isobar only if the condition

\begin{equation}
  \frac{eB}{2M_{\Delta}}\ll|M_{\Delta}-\left(M_{N}+m_{\pi}\right)|
\end{equation}
between the nucleon, $\Delta(1232)$ and pion masses is met. This
is a general situation for unstable particles and was discussed in
\cite{Ledwig(2010):nonAna} within a simpler field theory. In the
work \cite{Ledwig(2010):nonAna} only the one Feynman graph leading
to the above condition was investigated whereas we derive here the remaining two
conditions coming from the two more Feynman graphs. 

In the next section we will give necessary notations for the BFT and
will use them in the third section to investigate the self-energies
of stable and unstable particles. In the forth and fifth section we
apply the BFT to the nucleon and $\Delta\left(1232\right)$-isobar
and derive the above expressions. 

\section{Field equations in presence of an external em field}

\begin{figure}
  \includegraphics[scale=0.5]{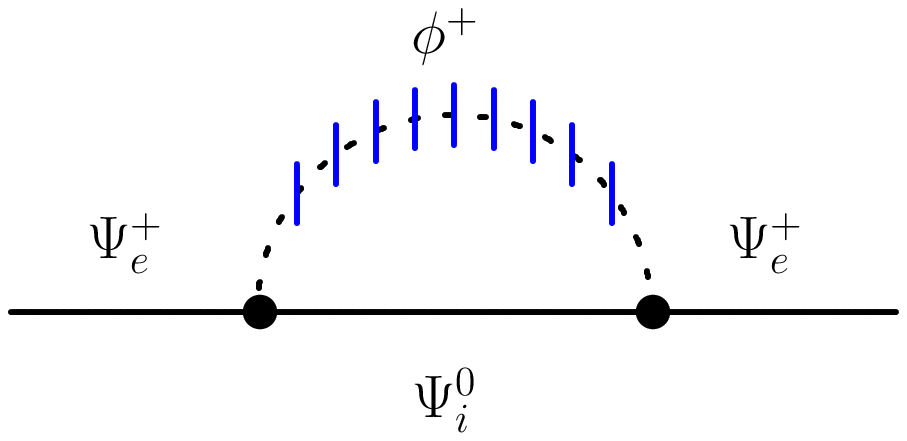}~~~\includegraphics[scale=0.5]{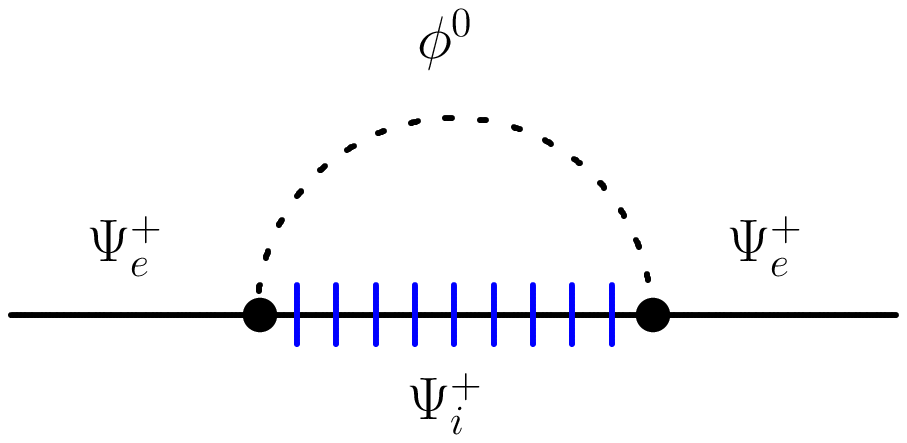}
  ~~~\includegraphics[scale=0.5]{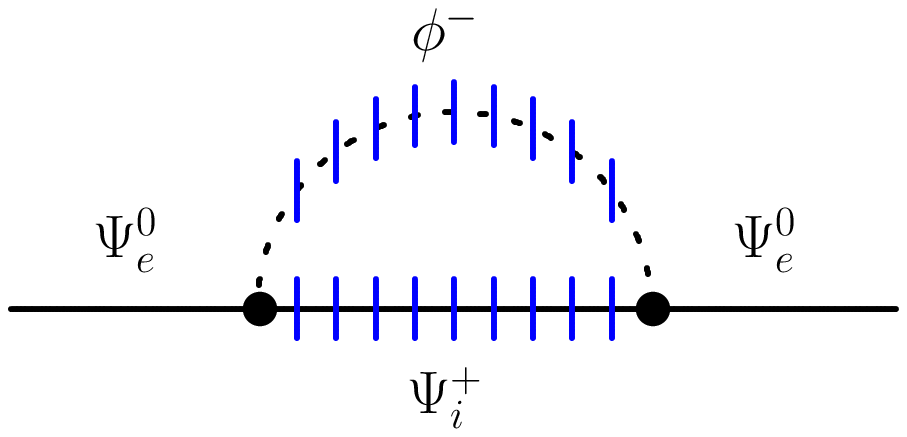}\\
  \includegraphics[scale=0.5]{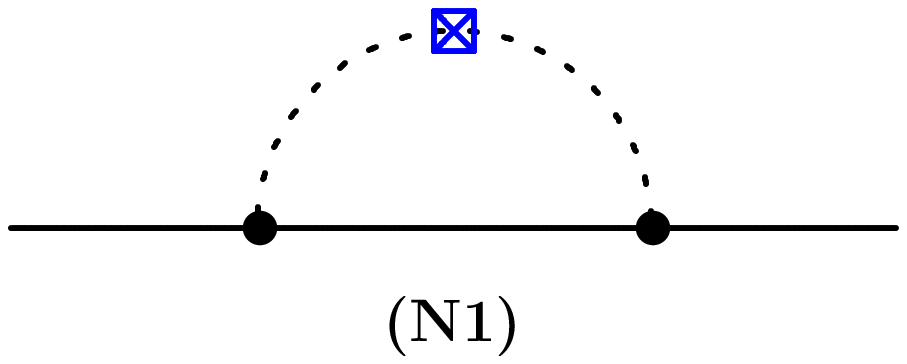}~~~\includegraphics[scale=0.5]{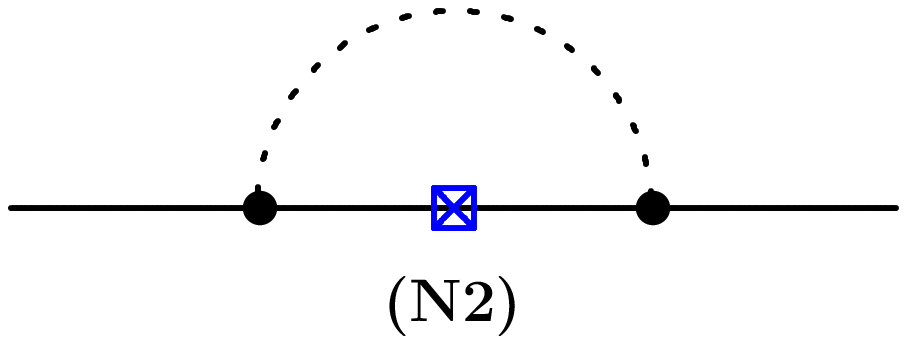}\\
  \includegraphics[scale=0.5]{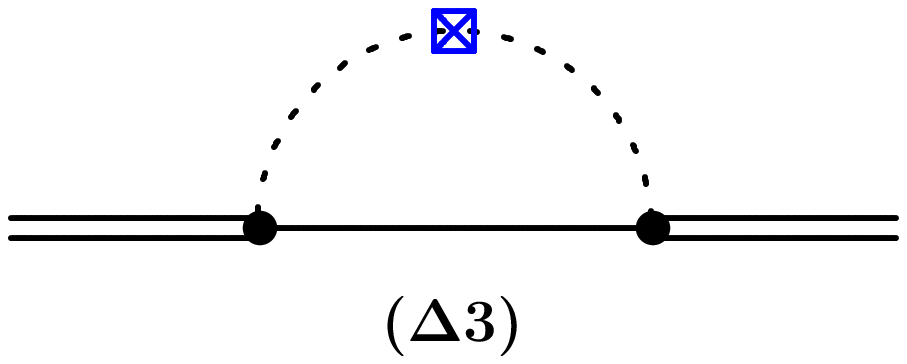}~~~\includegraphics[scale=0.5]{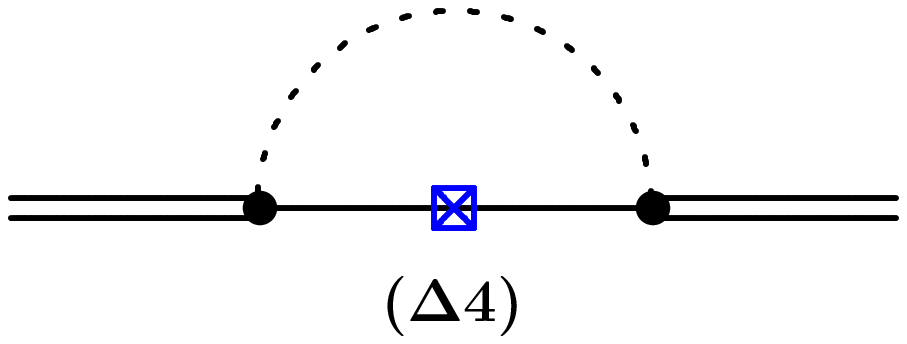}
  \caption{\label{fig:feynman_graphs} Upper row: Feynman graphs contributing
    to the self-energy in the presence of an external electromagnetic
    field. The blue lines indicate which loop-internal particle is affected
    by the field. Middle and lower row: Feynman graphs contributing to
    the three point function method to obtain the AMM. Single solid lines
    correspond to stable particles (e.g. the nucleon), double solid lines
    to unstable particles (e.g. the $\Delta\left(1232\right)$), dashed
    lines to (pseudo-) scalar particles (e.g. the pion) and the blue cross
    to the coupling of the photon.}
\end{figure}

We consider spin-1/2 and spin-3/2 fields moving in a constant electromagnetic
field given by the potential $A_{\mu}(x)=-\frac{1}{2}F_{\mu\nu}x^{\nu}$
with $F^{\mu\nu}=\partial^{[\mu}A^{\nu]}$ as the electromagnetic
field strength tensor. The Dirac equation for a particle $\Psi(x)$
of mass $M$ with the EM minimal substitution is:
\begin{equation}
  \left[\s\Pi-M\right]\Psi_{A}(x)=0\,\,\,,\label{eq:dirac_eq_A}
\end{equation}
where we take $e>0$ and define the momentum $\Pi_{\mu}=i\partial_{\mu}-eA_{\mu}(x)$.
This operator is non-commutative with 
\begin{equation}
  \left[\Pi_{\mu},\Pi_{\nu}\right]=\frac{1}{i}F_{\mu\nu}\,\,\,.\label{eq:commutation_relation}
\end{equation}
The spin $3/2$ field $\psi_{\mu}(x)$ satisfies the equation
\begin{equation}
  \left[\s\Pi-M\right]\Psi_{A}^{\mu}(x)=0,
\end{equation}
with the subsidiary conditions
\begin{eqnarray}
\gamma_{\mu}\Psi_{A}^{\mu}(x) & = & 0\,\,,\\
\Pi_{\mu}\Psi_{A}^{\mu}(x) & = & 0\,\,.
\end{eqnarray}
Because of the non-commutativity we have to symmetrize occurring expressions
and use for the propagators: 
\begin{equation}
  \frac{1}{\s\Pi-M}=\frac{1}{2}\left[\left(\s\Pi+M\right)\frac{1}{\Pi^{2}-M^{2}+\s
      F}+\frac{1}{\Pi^{2}-M^{2}+\s F}\left(\s\Pi+M\right)\right]\,\,,
\end{equation}
where we introduce the notation $\s F=\frac{1}{2i}\gamma^{\alpha}F_{\alpha\beta}\gamma^{\beta}$
\cite{Sommerfield}. With this we calculate the particle's self-energy
$\Sigma\left(\s F\right)$ and obtain its anomalous magnetic moment
(AMM) $\kappa$ through the linear energy shift:

\begin{equation}
  \langle\Psi_{A}|\Sigma\left(\s
  F\right)|\Psi_{A}\rangle=\langle\Sigma\left(0\right)\rangle-\langle\s
  F\rangle\frac{\kappa}{2M}+\mathcal{O}\left(\s F^{2}\right)\,\,\,.
\end{equation}
In Fig. \ref{fig:feynman_graphs} we list all Feynman graphs used
in this work for the BFT and three point function method. The upper
row shows all types of graphs that appear to the one-loop level in
the BFT, tadpole graphs not considered. 

For better reading we use the notation $\vec{B}=M^{2}\,\mathcal{B}\,\vec{e}_{z}$,
i.e. $\mathcal{B}=|\vec{B}|/M^{2}$, with which the linear approximation
to the self-energy reads:

\begin{eqnarray}
  \Sigma\left(B\right) & = &
  \Sigma\left(0\right)-M\frac{1}{2}\kappa\mathcal{B}
  +\mathcal{O}\left(B^{2}\right)\,\,\,,\label{eq:linear_approximation}\\
  \Sigma\left(-B\right)-\Sigma\left(B\right) & = &
  M\kappa\mathcal{B}+\mathcal{O}\left(B^{3}\right)\,\,\,,
\end{eqnarray}
where in the last equation the $B^{2}$ terms cancel out. The self-energy
formulas in this work omit some, but not all, $\mathcal{O}\left(B^{2}\right)$
terms. In particular non-analytic $B$ structures are preserved in
the\textbf{ }BFT\textbf{ }which are not in the one-photon approximation
Eq. (\ref{eq:linear_approximation}). 

\section{Scalar couplings and estimation of $\mathcal{O}(B^{2})$ effects}

We use the notations of the appendix and consider two spin 1/2 fields
$\Psi_{1}$, $\Psi_{2}$ interacting with a scalar field $\phi$:

\begin{eqnarray}
  \mathcal{L} & = & \sum_{a=1}^{2}\,\overline{\Psi_{a}}\left(i\s
  D_{a}-M_{a}\right)\Psi_{a}+\frac{1}{2}\left(D_{\mu}\phi\right)\left(D^{\mu}\phi\right)
  -\frac{1}{2}m^{2}\phi^{2}+\sum_{a,b=1}^{2}g\overline{\Psi_{a}}\Gamma_{ab}\Psi_{b}\phi\,\,\,,\label{eq:ps_lagrangian}
\end{eqnarray}
with the covariant derivative $D_{\mu}=\partial_{\mu}+iqeA_{\mu}$,
$q$ as the charge of the field with $e>0$ and $\Gamma_{ab}$ either
$1$ or $\gamma_{5}$. In the upper row of Fig. \ref{fig:feynman_graphs}
we show all the types of self-energy graphs that can occur for a charged
or uncharged external particle $\Psi_{ex}$. The corresponding expressions
$\Sigma_{i}\left(B\right)$ with a loop-internal particle $\Psi_{in}$
are:

\begin{eqnarray}
  \langle\Psi_{ex}|\Sigma_{1}\left(\s F\right)|\Psi_{ex}\rangle & = &
  \frac{g^{2}}{i}
  \langle\Psi_{ex}|\int\tilde{dl}\Gamma_{1}\frac{1}{\left[\s
      l-M_{in}+i\varepsilon\right]}
  \Gamma_{1}\frac{1}{\left[\left(\Pi-l\right)^{2}-m^{2}+i\varepsilon\right]}
  |\Psi_{ex}\rangle\,\,\,,\\
  \langle\Psi_{ex}|\Sigma_{2}\left(\s F\right)|\Psi_{ex}\rangle & = & 
  \frac{g^{2}}{i}\langle\Psi_{ex}|\int\tilde{dl}\Gamma_{2}\frac{1}{\left[\s\Pi-\s
      l-M_{in}+i\varepsilon\right]}
  \Gamma_{2}\frac{1}{\left[l^{2}-m^{2}+i\varepsilon\right]}|\Psi_{ex}\rangle\,\,\,,\\
  \langle\Psi_{ex}|\Sigma_{3}\left(\s F\right)|\Psi_{ex}\rangle & = &
  \frac{g^{2}}{i}\langle\Psi_{ex}|\int\tilde{dl}\Gamma_{3}\frac{1}{\left[\s\Pi-\s
      l-M_{in}+i\varepsilon\right]}\Gamma_{3}\frac{1}{\left[\left(\Pi_{+}
      +l\right)^{2}-m^{2}+i\varepsilon\right]}|\Psi_{ex}\rangle\,\,\,,
\end{eqnarray}
with $\Pi_{+}^{\mu}=i\partial^{\mu}+eA^{\mu}(x)$. For a third-spin
projection of $+1/2$ we can write these expressions in dimensional
regularization as: 

\begin{eqnarray}
  \Sigma_{i} & = &
  \frac{-g^{2}}{\left(4\pi\right)^{2}}M_{ex}\int_{-\alpha_{i}}^{1-\alpha_{i}}dz\left[s_{5}
    \left(z+\alpha_{i}\right)+r\right]\left[L+\ln\left(z^{2}-\lambda_{i}^{2}
    -i\varepsilon\right)+\left(\delta_{i1}+\delta_{i2}\right)
    \ln\left(1-\mathcal{B}\right)\right]
  +\mathcal{O}\left(\mathcal{B}^{2}\right)\,\,\,,\label{eq:self_energy_ps}
\end{eqnarray}
with $\alpha_{i}$, $\lambda_{i}$ given below and $s_{5}=-1$ for
$\Gamma_{i}=\gamma_{5}$ and $+1$ for $\Gamma_{i}=1$. We see that
for $\vec{B}=0$ these expressions are the same and for $\vec{B}\neq0$
several different logarithms occur. The formula Eq. (\ref{eq:self_energy_ps})
omits some, but not all, $B^{2}$ terms and the integrated solution
for the real and imaginary parts are:

\begin{eqnarray}
  \mbox{Re}\,\Sigma_{i} & = & \frac{-g^{2}}{\left(4\pi\right)^{2}}M_{ex}\,\,
  \Big[+\left(s_{5}\alpha_{i}+r\right)\left(\beta_{i}\ln\left(\beta_{i}^{2}
    -\lambda_{i}^{2}\right)+\alpha_{i}\ln\left(\alpha_{i}^{2}-\lambda_{i}^{2}\right)-2\right)\nn\\
    &  &
    +s_{5}\frac{1}{2}\left(\alpha_{i}^{2}-\beta_{i}^{2}+\left(\beta_{i}^{2}
    -\lambda_{i}^{2}\right)\ln\left(\beta_{i}^{2}-\lambda_{i}^{2}\right)
    -\left(\alpha_{i}^{2}-\lambda_{i}^{2}\right)\ln\left(\alpha_{i}^{2}
    -\lambda_{i}^{2}\right)\right)\nn\\
    &  &
    +\left(\delta_{i1}+\delta_{i2}\right)\ln\left(1-\mathcal{B}\right)\left(s_{5}\alpha_{i}
    +r+\frac{s_{5}}{2}\left(\beta_{i}^{2}-\alpha_{i}^{2}\right)\right)
    +\left(s_{5}\alpha_{i}+r\right)\Omega_{i}\,\,\Big]\,\,\,,\\
  \mbox{Im}\,\Sigma_{i} & = &
  \frac{g^{2}\pi}{\left(4\pi\right)^{2}}M_{ex}\left(s_{5}\alpha_{i}+r\right)2\lambda_{i}
\end{eqnarray}
with $\beta_{i}=1-\alpha_{i}$ and 

\begin{eqnarray}
  \Omega_{i} & = & \begin{cases}
    \begin{array}{c}
      2\sqrt{-\lambda_{i}^{2}}\left(\arctan\frac{\beta_{i}}{\sqrt{-\lambda_{i}^{2}}}
      +\arctan\frac{\alpha_{i}}{\sqrt{-\lambda_{i}^{2}}}\right)\\
      2\sqrt{\lambda_{i}^{2}}\left(\mbox{arctanh}\frac{\beta_{i}}{\sqrt{\lambda_{i}^{2}}}
      +\mbox{arctanh}\frac{\alpha_{i}}{\sqrt{\lambda_{i}^{2}}}\right)\end{array}
    &     \begin{array}{c}
      \lambda_{i}^{2}<0\\
      \lambda_{i}^{2}>0\end{array}\end{cases}\,\,\,,\label{eq:omega_contributions}
\end{eqnarray}

\begin{eqnarray}
  \alpha_{0}=\frac{1}{2}\left(1+r^{2}-\mu^{2}\right) & \,\,,\,\, & 
  \lambda_{0}^{2}=\alpha_{0}^{2}-r^{2}\,\,,\\
  \alpha_{1}=\frac{1}{2\left(1-\mathcal{B}\right)}\left(1+r^{2}-\mu^{2}-\mathcal{B}\right)
  & \,\,,
  \,\, & \lambda_{1}^{2}=\alpha_{1}^{2}-\frac{r^{2}}{1-\mathcal{B}}\,\,,\\
  \alpha_{2}=\frac{1}{2\left(1-\mathcal{B}\right)}\left(1+r^{2}-\mu^{2}-2\mathcal{B}\right)
  & \,\,
  ,\,\, & \lambda_{2}^{2}=\alpha_{2}^{2}-\frac{r^{2}-\mathcal{B}}{1-\mathcal{B}}\,\,,\\
  \alpha_{3}=\frac{1}{2}\left(1+r^{2}-\mu^{2}-\mathcal{B}\right) & \,\,,\,\, &
  \lambda_{3}^{2}=\alpha_{3}^{2}-r^{2}+\mathcal{B}\,\,.
\end{eqnarray}
The self-energies obtain imaginary parts if $\lambda_{i}^{2}$ becomes
positive together with $-\lambda_{i}<\beta_{i}$ and/or $\lambda_{i}>-\alpha_{i}$.
In addition, the non-analytic contributions $\Omega_{i}$ give constrains
on when the self-energies can be expanded for small $\mathcal{B}$.
These constrains can generically be written as:
\begin{eqnarray}
  \sqrt{f_{i}(\mu,r)\mathcal{B}+\lambda_{0}^{2}} & \to &
  |\mathcal{B}|<|\frac{\lambda_{0}^{2}}{f_{i}(\mu,r)}|\,\,\,,
\end{eqnarray}
for a certain function $f_{i}$ depending on the type of graph. One
of these constrains, coming from the first graph in Fig. \ref{fig:feynman_graphs},
was investigated in \cite{Ledwig(2010):nonAna} for the situation
of $M_{ex}>M_{in}$. In the following we investigate the remaining
two as well as further applications of Eq. (\ref{eq:self_energy_ps}).

\subsubsection{Nucleon-pion system}

The first example is the nucleon-pion ($N$, $\pi$) system with pseudo-scalar
couplings, $\Gamma_{a}^{NN\pi}=g_{a}\gamma_{5}$. For this we have
$s_{5}=-1$, $r=1$, $M_{ex}=M_{N}$, $m=m_{\pi}$ and write for the
proton and neutron energies:

\begin{eqnarray}
  \Sigma_{p}\left(B\right) & = & 2\Sigma_{1}\left(B\right)+\Sigma_{2}\left(B\right)\,\,\,,\label{eq:ps_proton}\\
  \Sigma_{n}\left(B\right) & = &
  \Sigma_{3}\left(0\right)+2\Sigma_{3}\left(B\right)\,\,\,,\label{eq:ps_neutron_s3}
\end{eqnarray}
with
\begin{eqnarray}
  \Sigma_{i} & = &
  \frac{-g^{2}}{\left(4\pi\right)^{2}}M_{N}\int_{0}^{1}dz\,\left(1-z\right)\left[L+\ln\left(z\mu^{2}
    +\left(1-z\right)^{2}+\mathcal{B}_{i}-i\varepsilon\right)\right]\,\,\,,
\end{eqnarray}
and $\mathcal{B}_{1}=z\left(1-z\right)\mathcal{B}$, $\mathcal{B}_{2}=-\left(1-z\right)^{2}\mathcal{B}$
and $\mathcal{B}_{3}=-\left(1-z\right)\mathcal{B}$. It is easy to
check that we get from this form the same AMM as obtained from the
usual three point function method:
\begin{eqnarray}
  \kappa_{p}=2\kappa_{1}+\kappa_{2} & \,\,\,\,\,,\,\,\,\,\, & \kappa_{n}=-2\kappa_{1}+2\kappa_{2}\,\,\,,\\
  \kappa_{1}=\frac{g^{2}}{\left(4\pi\right)^{2}}\int_{0}^{1}dz\frac{2z\left(1-z\right)^{2}}{z\mu^{2}+\left(1-z\right)^{2}}
  & \,\,\,\,\,,\,\,\,\,\, &
  \kappa_{2}=\frac{g^{2}}{\left(4\pi\right)^{2}}\int_{0}^{1}dz\frac{-2\left(1-z\right)^{3}}{z\mu^{2}+\left(1-z\right)^{2}}\,\,\,.
\end{eqnarray}
According to this, we can also write for the neutron and iso-vector
self-energies:
\begin{eqnarray}
  \Sigma_{n}\left(B\right) & = & 3\Sigma_{1}\left(0\right)-2\Sigma_{1}\left(B\right)+2\Sigma_{2}\left(B\right)\,\,\,,\label{eq:ps_neutron_s12}\\
  \Sigma_{v}\left(B\right) & = &
  4\Sigma_{1}\left(B\right)-\Sigma_{2}\left(B\right)\,\,\,,
\end{eqnarray}
where the difference of Eq.(\ref{eq:ps_neutron_s12}) to Eq.(\ref{eq:ps_neutron_s3})
is of order $B^{2}$. 

\begin{figure}
  \caption{\label{fig:ps_self_energies}Energy shift of the proton (left) and
    neutron (right) with pseudo-scalar coupling for $m_{\pi}=139$ MeV.
    The upper row corresponds to the shift with $\Sigma\left(B\right)-\Sigma\left(0\right)$
    and the lower row to $\Sigma\left(-B\right)-\Sigma\left(B\right)$.
    The linear solid green lines corresponds to the linear approximation
    Eq. (\ref{eq:linear_approximation}) while the curved solid red lines
    to Eqs. (\ref{eq:ps_proton},\ref{eq:ps_neutron_s3}). The solid blue
    line shows the result of Eq. (\ref{eq:ps_neutron_s12}). The dotted
    lines correspond to the the imaginary parts.}
  \includegraphics[scale=0.4]{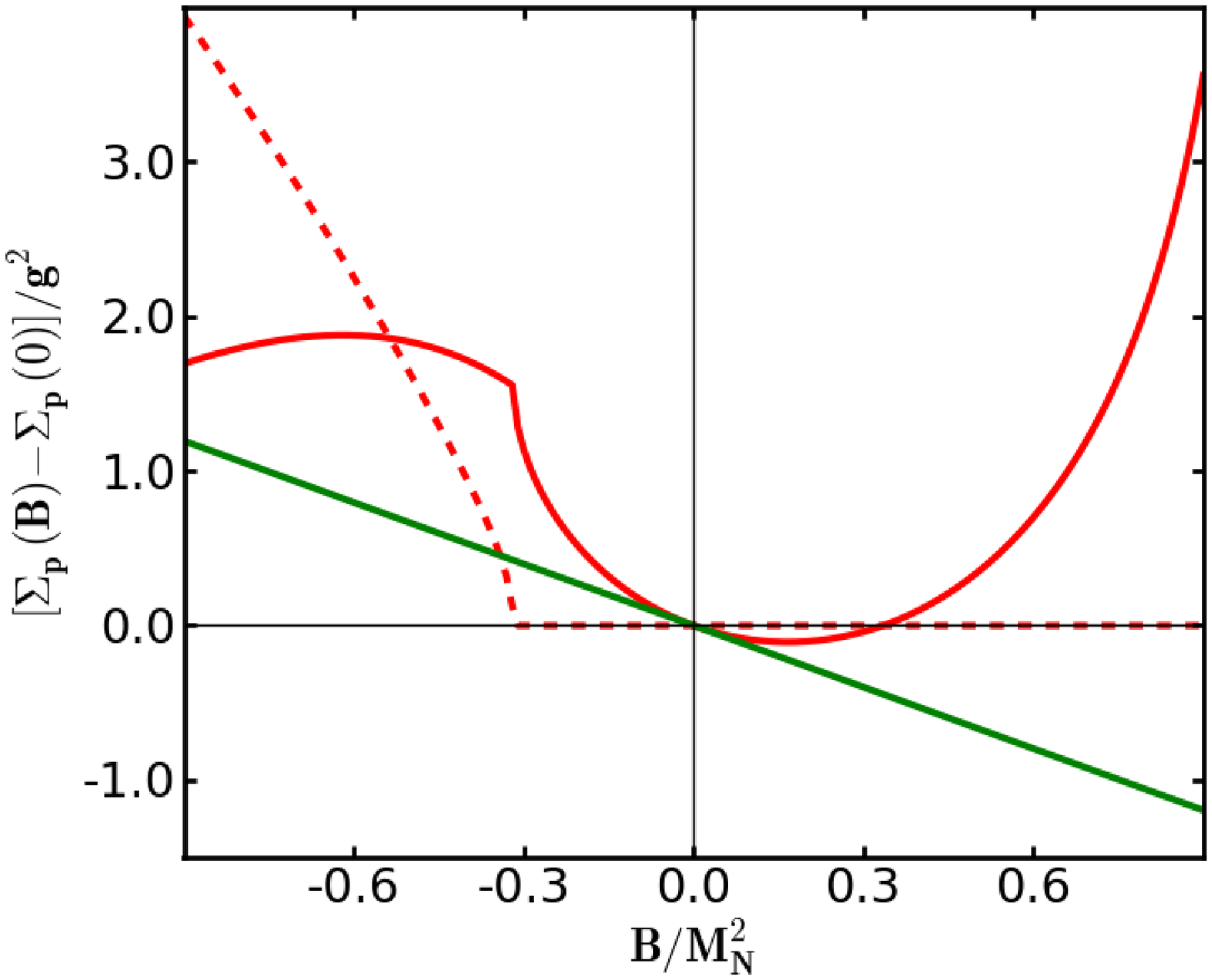}~~~~~~~~\includegraphics[scale=0.4]{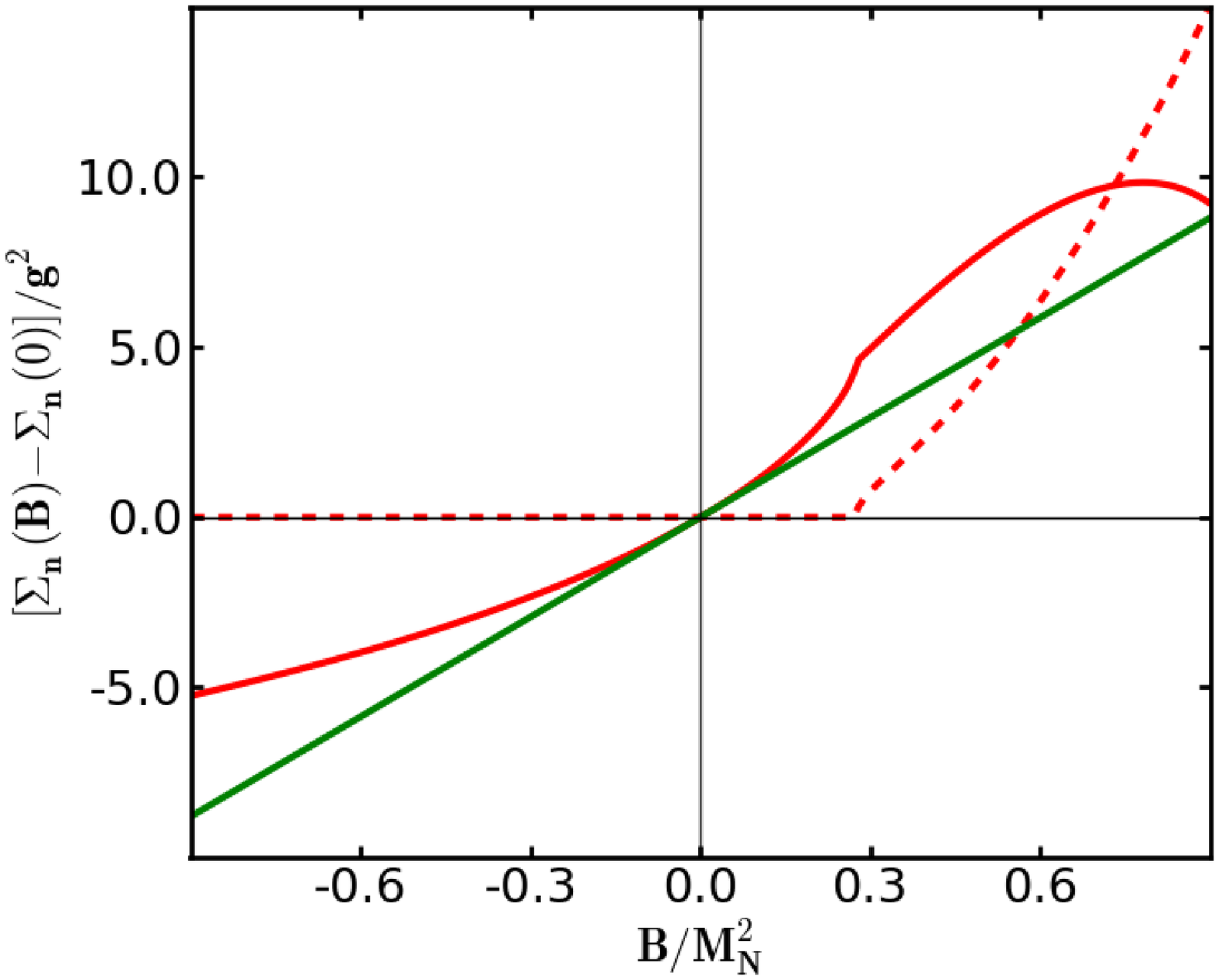}\\
  \includegraphics[scale=0.4]{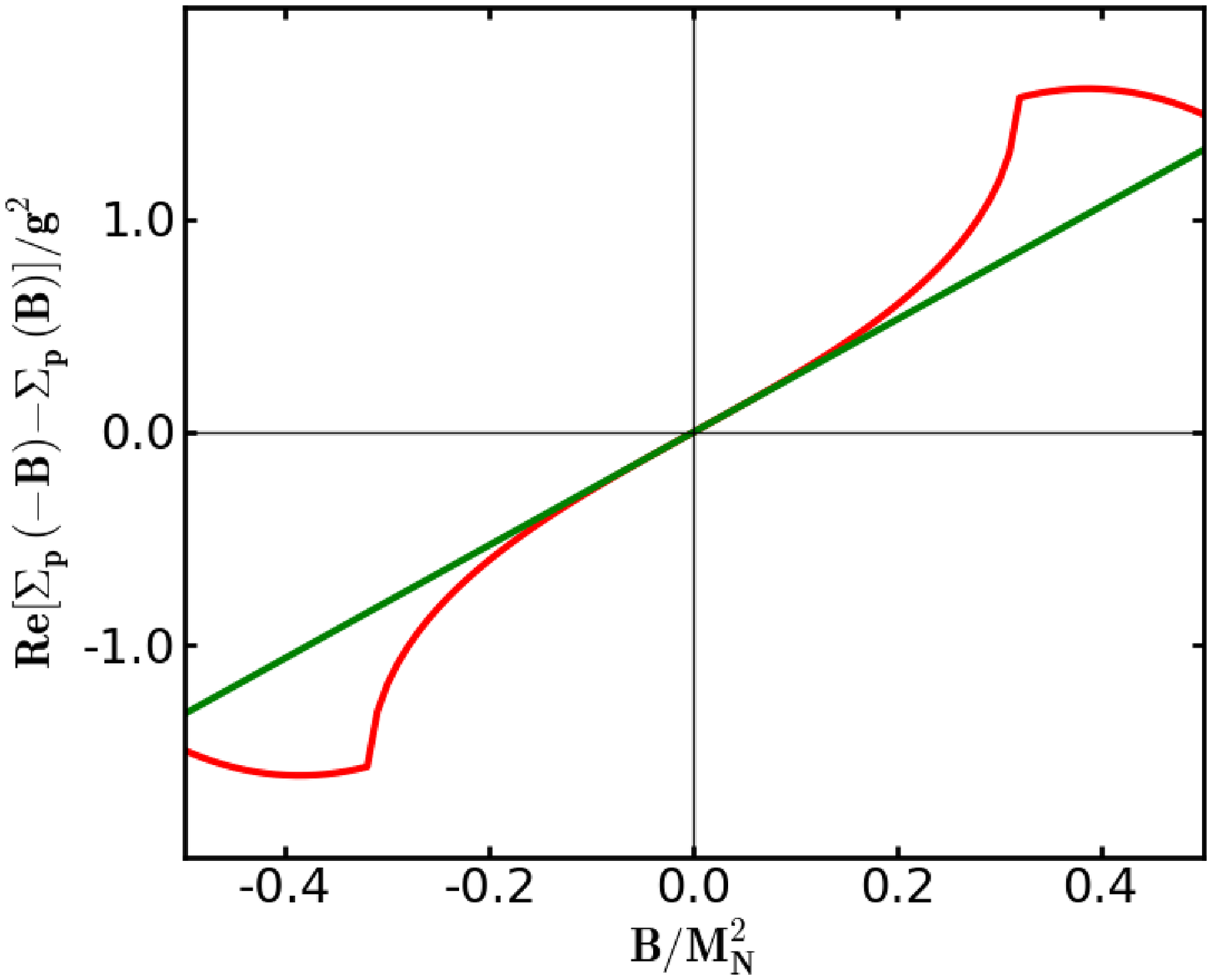}~~~~~~~~\includegraphics[scale=0.4]{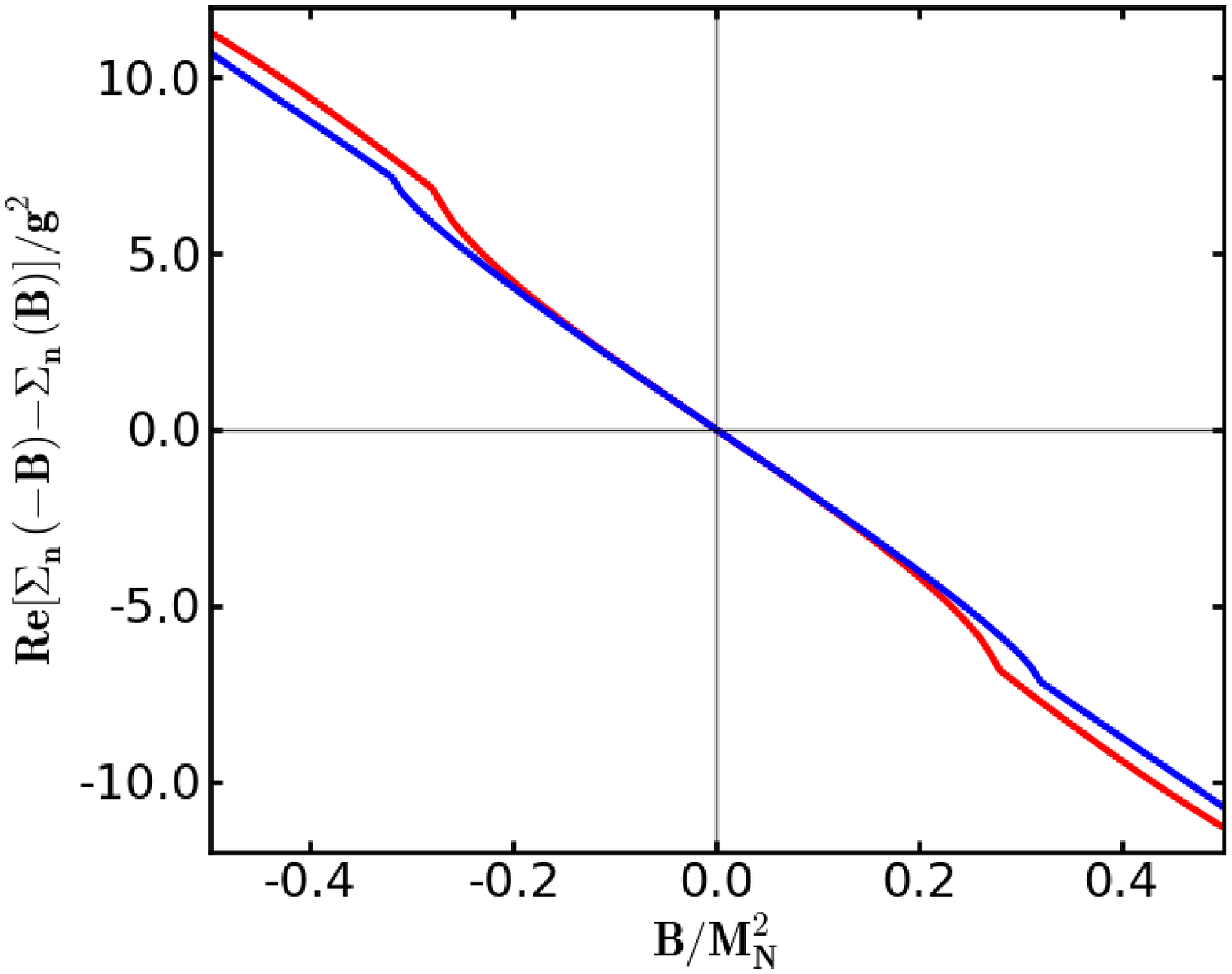}
\end{figure}

In Fig. \ref{fig:ps_self_energies} we plot the nucleon energy shifts
together with the linear approximation with $M_{N}=939$ MeV and $m_{\pi}=139$
MeV. We see that the signs of the AMM are in agreement with phenomenology,
i.e. $\kappa_{p}>0$ and $\kappa_{n}<0$. To estimate $\mathcal{O}\left(B^{2}\right)$
effects, we use the difference of the red and blue lines in the lower
right neutron graph. The combination $\Sigma\left(-B\right)-\Sigma\left(B\right)$
does not have $\mathcal{O}\left(B^{2}\right)$ contributions whereas
the difference of the expressions Eq. (\ref{eq:ps_neutron_s3}) and
Eq. (\ref{eq:ps_neutron_s12}) is of $\mathcal{O}\left(B^{2}\right)$.
Plotting the same results for various pion masses shows that the $B^{2}$
effects are small for magnetic field strengths of $|B|<\frac{1}{5}M_{N}^{2}$
for pion masses larger than $m_{\pi}=100$ MeV and for $|B|<\frac{1}{2}M_{N}^{2}$
with pion masses around $m_{\pi}=600$ MeV. Within these region the
Eqs.(\ref{eq:linear_approximation},\ref{eq:ps_neutron_s3},\ref{eq:ps_neutron_s12})
give approximately the same results. We estimate therefore that the
self-energy formulas are applicable for magnetic field strengths of
$|B|<\frac{1}{5}M_{N}^{2}$ with pion masses between the physical
point $m_{\pi}=140$ MeV up to $m_{\pi}=600$ MeV. This is the applied
pion mass range in lattice QCD calculations.

In addition we have also for the nucleon the non-analytic $\mathcal{B}$
expressions, Eq. (\ref{eq:omega_contributions}), which would constrain
a definition of the magnetic moment by the linear energy shift. These
constrains read:

\begin{eqnarray}
  \Sigma_{1,3}: &  & |\mathcal{B}|<|\mu\left(\mu-2\right)|\,\,\,,\\
  \Sigma_{1,3}: &  & |\mathcal{B}|<|\mu\left(\mu+2\right)|\,\,\,,\\
  \Sigma_{2}: &  & |\mathcal{B}|<|\frac{1}{4}\mu^{2}-1|\,\,\,,
\end{eqnarray}
which can always be fulfilled for $0<\mu<2$. 

Further, we obtain for the nucleon self-energy an imaginary part when
a stronger magnetic field is applied. The imaginary parts come only
from $\Sigma_{1}$ and $\Sigma_{3}$ and read:

\begin{eqnarray}
  \mbox{Im}\Sigma_{i} & = &
  M_{N}\frac{g^{2}\pi}{\left(4\pi\right)^{2}}\left(1-\alpha_{i}\right)2\lambda_{i}\,\,\,,
\end{eqnarray}
for $\mathcal{B}\leq-\mu\left(2+\mu\right)$ in $\Sigma_{1}$ and
$\mathcal{B}\ge\mu\left(2-\mu\right)$ in $\Sigma_{3}$. 

The nucleon results of this section are obtained from the general
expression Eq. (\ref{eq:self_energy_ps}) for particles with Yukawa
couplings as in the Lagrangian Eq. (\ref{eq:ps_lagrangian}) and are
transcript-able for other stable particles. 

\subsubsection{Nucleon-pion-resonance system}

For the second example we include a resonance ($R$) of mass $M_{R}$
with the coupling $\Gamma_{a}^{NR\pi}=ig_{a}$ and get with $s_{5}=1$
and $r=M_{N}/M_{R}$ the self-energies: 

\begin{eqnarray}
  \Sigma_{i} & = &
  \frac{g^{2}}{\left(4\pi\right)^{2}}M_{R}\int_{0}^{1}dz\,\left(z+r\right)\left[L+\ln\left(z\mu^{2}
    -z\left(1-z\right)+\left(1-z\right)r^{2}+\mathcal{B}_{i}-i\varepsilon\right)\right]
  +\mathcal{O}\left(B^{2}\right)\,\,\,.\label{eq:resonance_self_energy}
\end{eqnarray}
From this, we also recover the results of the three point method:
\begin{equation}
  \kappa_{1}=\frac{g^{2}}{\left(4\pi\right)^{2}}\int_{0}^{1}dz\,\frac{2z\left(1-z\right)
    \left(-z-r\right)}{z\mu^{2}-z\left(1-z\right)+\left(1-z\right)r^{2}-i\varepsilon}\,\,\,\,\,\,\,\,\,,
  \,\,\,\,\,\,\,\,\,\kappa_{2}=\frac{g^{2}}{\left(4\pi\right)^{2}}\int_{0}^{1}dz\,\frac{2\left(1-z\right)^{2}
    \left(z+r\right)}{z\mu^{2}-z\left(1-z\right)+\left(1-z\right)r^{2}-i\varepsilon}\,\,\,.
\end{equation}
Since the resonance is unstable we have the following imaginary parts
for the self-energies:
\begin{eqnarray}
\mbox{Im}\Sigma_{i} & = &
-M_{R}\frac{g^{2}}{\left(4\pi\right)^{2}}\pi\left(r+\alpha_{i}\right)2\lambda_{i}\label{eq:imag_resonance}
\end{eqnarray}
for $\alpha_{i}>\lambda_{i}$. Explicitly, these parts are present
for magnetic fields of 
\begin{eqnarray}
  \Sigma_{1}: &  & \mathcal{B}<-\sqrt{\left(1-r^{2}-\mu^{2}\right)^{2}-4\lambda_{0}^{2}}+1-r^{2}-\mu^{2}\,\,\,,\label{eq:imag_sigma1_resonance}\\
  \Sigma_{2}: &  & \mathcal{B}>-\frac{\lambda_{0}^{2}}{\mu^{2}}\,\,\,,\label{eq:imag_sigma2_resonance}\\
  \Sigma_{3}: &  &
  \mathcal{B}>+\sqrt{\left(1-r^{2}+\mu^{2}\right)^{2}-4\lambda_{0}^{2}}-1+r^{2}+\mu^{2}\,\,\,.\label{eq:imag_sigma3_resonance}
\end{eqnarray}
In the case of $\Sigma_{2}$ and $\Sigma_{3}$ with a magnetic field
of $\mathcal{B}\geq r^{2}$ we have $\lambda_{i}>\alpha_{i}$, Eq.
(\ref{eq:imag_resonance}) has do be altered accordingly and an additional
cusp is present at $\mathcal{B}=r^{2}$.

\begin{figure}
  \caption{\label{fig:sigma_resonance}Self-energies of a resonance in an external
    magnetic field $B$. The linear green lines correspond to the linear
    approximation Eq. (\ref{eq:linear_approximation}) while the curved
    red lines to Eqs. (\ref{eq:resonance_self_energy}). The solid lines
    show the real parts and the dotted lines the imaginary parts.}
  \includegraphics[scale=0.27]{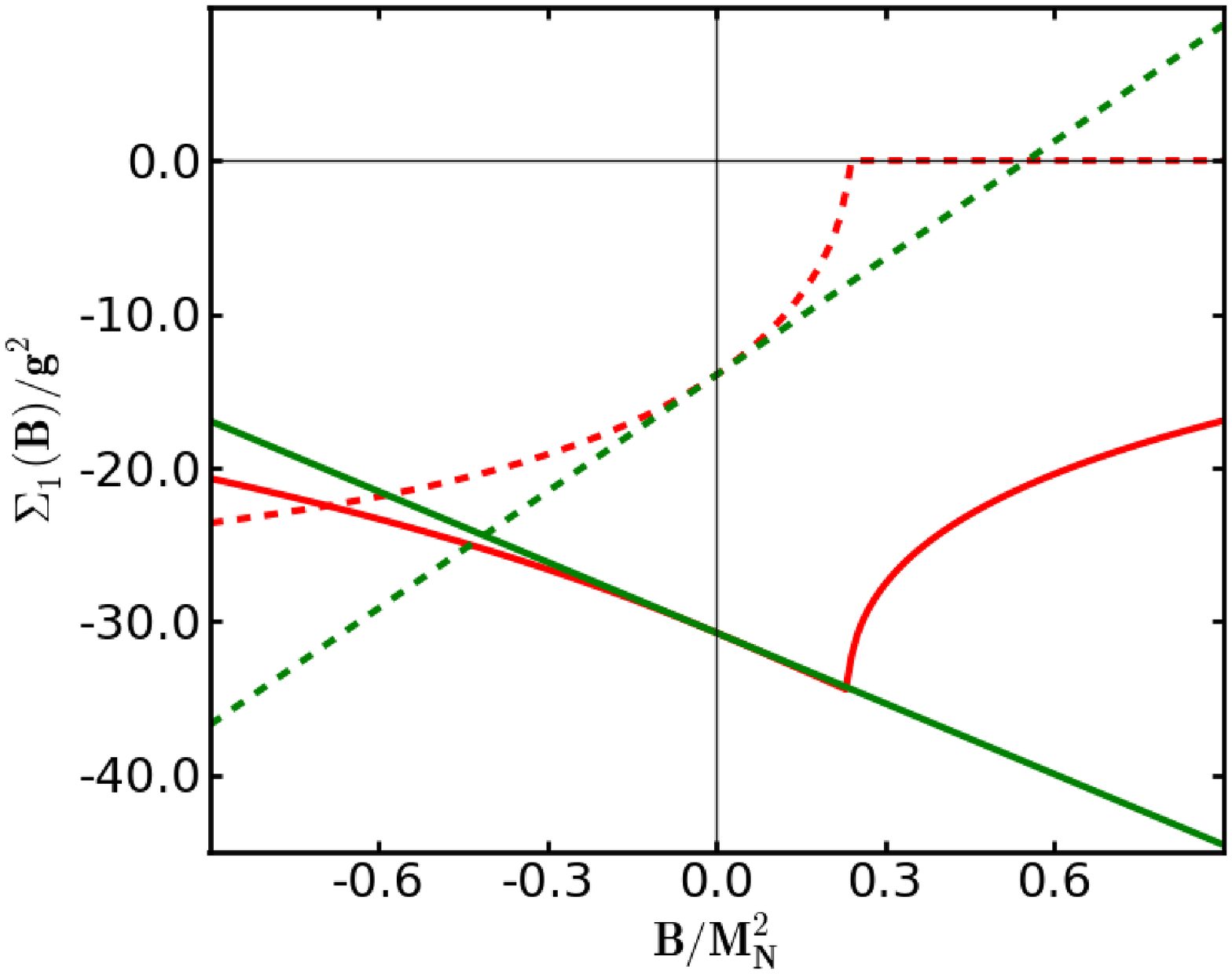}~~~~\includegraphics[scale=0.27]{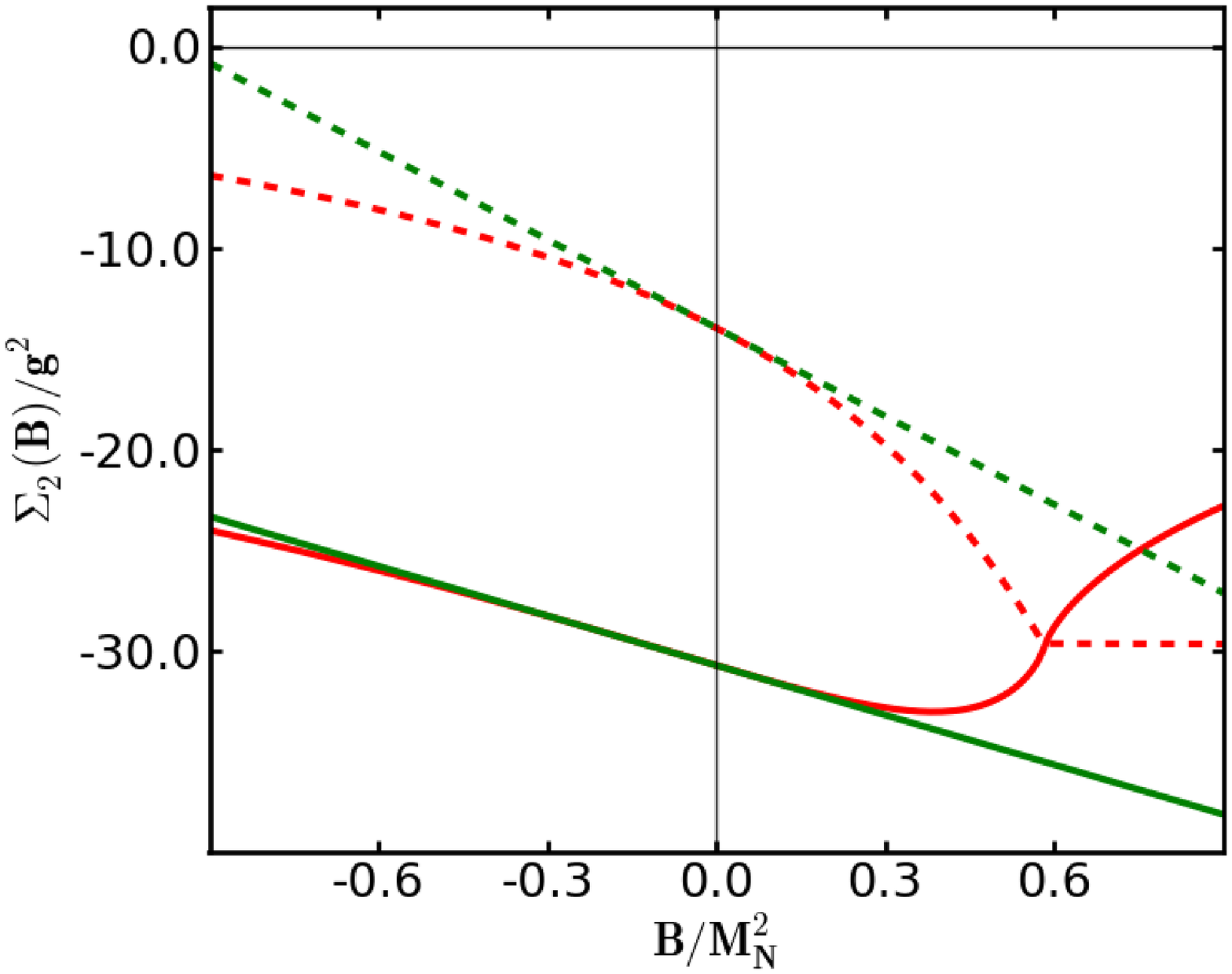}~~~~\includegraphics[scale=0.27]{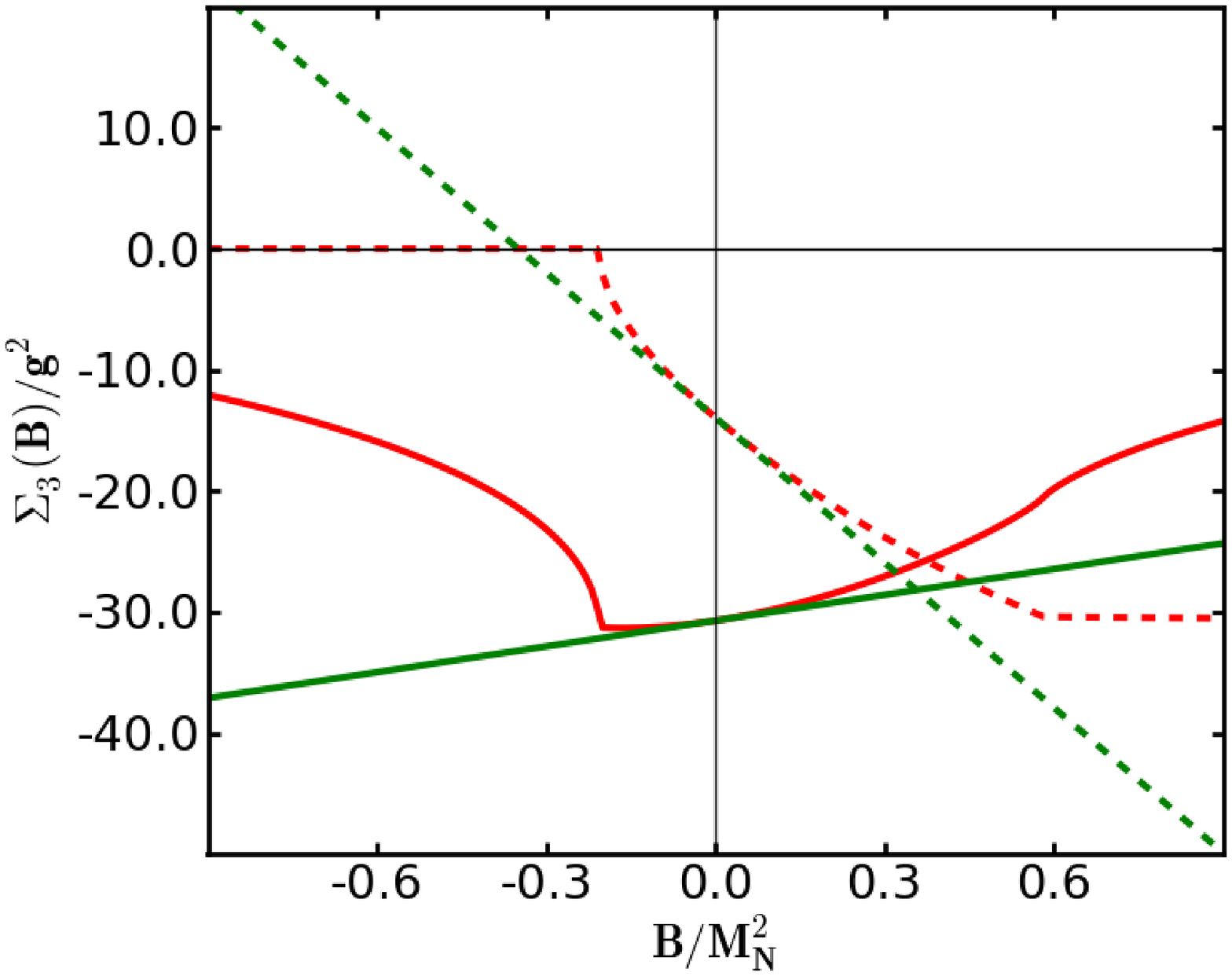}
\end{figure}

In Fig. \ref{fig:sigma_resonance} we show all three possible self-energies
for the parameters $M_{N}=939$ MeV, $M_{R}=1232$ MeV and $m_{\pi}=139$
MeV together with the linear approximations. For the graph $\Sigma_{1}$
we see the linear behavior near $B=0$ and a cusp appearing according
to Eq. (\ref{eq:imag_sigma1_resonance}). This graph was investigated
in \cite{Ledwig(2010):nonAna}. In the case of $\Sigma_{2}$ we see
only one cusp at $\mathcal{B}=r^{2}\approx0.6$ since for the present
mass constellation Eq. (\ref{eq:imag_sigma2_resonance}) is fulfilled
for all $|\mathcal{B}|<1$ and an imaginary part is steadily present.
However, choosing $m_{\pi}$ closer to the mass-gap $M_{R}-M_{N}$
the second cusp appears on the $\mathcal{B}<0$ side. The occurrence
of such two cusps can be seen for $\Sigma_{3}$. The cusp for $\mathcal{B}<0$
is due to the imaginary part from Eq. (\ref{eq:imag_sigma3_resonance})
and the one for $\mathcal{B}>0$ due to $\mathcal{B}\geq r^{2}$.
As we approach with $m_{\pi}$ the mass-gap $M_{R}-M_{N}$, i.e. $\mu=1-r$,
all cusps corresponding to Eqs. (\ref{eq:imag_sigma1_resonance},\ref{eq:imag_sigma2_resonance},\ref{eq:imag_sigma3_resonance})
converge on $B=0$ where we have $1-r-\mu=0$. For a $m_{\pi}$ mass
larger than the mass gap the resonance will not be unstable anymore
and we get similar results as in the previous example. 

The explicit conditions to expand the self-energies for small $\mathcal{B}$
due to the non-analytic contributions $\Omega_{i}$, Eq. (\ref{eq:omega_contributions}),
are:
\begin{eqnarray}
  \Sigma_{1}: &  & |\mathcal{B}|<2|1-r-\mu|\,\,\,,\\
  \Sigma_{2}: &  & |\mathcal{B}|<\frac{2r}{1-r}|1-r-\mu|\,\,\,,\\
  \Sigma_{3}: &  & |\mathcal{B}|<2r|1-r-\mu|\,\,\,,
\end{eqnarray}
for the parameters $r<1$, $\mu>0$ and $\mu<1+r$. The first condition
was found in \cite{Ledwig(2010):nonAna}. Which condition is the
most strict one on $\mathcal{B}$ depends on the charge of the external
particle and the actual values of the participating masses. For $\mu<r$
and $r>1/2$ it is the first one. 

\section{Nucleon self-energy }

We apply now the BFT to a more involved situation, namely, the nucleon
self-energy in the SU(2) chiral perturbation theory of \cite{BChPTLagrangian}:

\begin{eqnarray}
  \mathcal{L}_{N\pi} & = & \overline{N}(i\s
  D-M_{N})N-\frac{g_{A}}{2f_{\pi}}\overline{N}\tau^{a}\left(\s
  D^{ab}\pi^{b}\right)\gamma_{5}N+\frac{1}{2}(D_{\mu}^{ab}\pi^{b})(D_{ac}^{\mu}\pi^{c})
  -\frac{1}{2}m_{\pi}^{2}\pi_{a}\pi^{a}\,\,\,.
\end{eqnarray}
The covariant derivatives are given in the appendix and the nucleon
axial-vector and the pion decay constants are $g_{A}=1.27$ and $f_{\pi}=92.4$
MeV. We consider the two graphs $\Sigma_{1}$ and $\Sigma_{2}$ in
Fig. \ref{fig:feynman_graphs} and obtain for the nucleon the following
unrenormalized self-energies in $d=4-2\varepsilon$ dimensions:

\begin{eqnarray}
  \Sigma_{1}^{N}\left(B\right) & = &
  i\left(\frac{g_{A}}{2f_{\pi}}\right)^{2}M_{N}
  \int_{0}^{1}dz\Big[-M_{N}^{2}\left(1-z\right)^{3}J_{2}+\left(-6+3z\right)J_{1}
    +\left(3-z\right)\varepsilon J_{1}+Bz^{2}\left(3-z\right)J_{2}\Big]\,\,\,,\\
  \Sigma_{2}^{N}\left(B\right) & = &
  i\left(\frac{g_{A}}{2f_{\pi}}\right)^{2}M_{N}
  \int_{0}^{1}dz\Big[-M_{N}^{2}\left(1-z\right)^{3}J_{2}+\left(-6+3z\right)J_{1}+\left(3-z\right)\varepsilon J_{1}\nn\\
    &  &
    \,\,\,\,\,\,\,\,\,\,\,\,\,\,\,\,\,\,\,\,\,\,\,\,\,\,\,\,\,\,\,\,\,\,\,\,\,\,\,\,\,\,\,\,\,\,\,\,\,\,
    +B\left(2\left(1-z\right)^{4}M_{N}^{2}J_{3}+\left(3-8z+6z^{2}-z^{3}\right)J_{2}-\left(3-4z+z^{2}\right)\varepsilon
    J_{2}\right)\Big]\,\,\,.
\end{eqnarray}
The loop integrals $J_{i}=J_{i}\left(\mathcal{M}_{N}\right)$ with
$\mathcal{M}_{N}=zm_{\pi}^{2}+\left(1-z\right)^{2}M_{N}^{2}+z\left(1-z\right)B$
are listed in the appendix. The expressions $\Sigma_{1}$ and $\Sigma_{2}$
only differ by $B$ dependent terms and we write for the nucleon self-energies:
\begin{eqnarray}
  \Sigma_{p}\left(B\right) & = & 2\Sigma_{1}\left(B\right)+\Sigma_{2}\left(B\right)\,\,\,,\\
  \Sigma_{n}\left(B\right) & = &
  3\Sigma_{1}\left(0\right)-2\Sigma_{1}\left(B\right)+2\Sigma_{2}\left(B\right)\,\,\,.
\end{eqnarray}
By integrating the Feynman parameter we get the $\tilde{MS}$ renormalized
proton self-energy
\begin{eqnarray}
  \Sigma_{p}\left(\mathcal{B}\right) & = &
  \frac{M_{N}C_{N}}{24\left(1-\mathcal{B}\right)^{4}}
  \left[n_{1}+n_{2}\ln\mu+\frac{n_{3}}{\sqrt{4\mu^{2}-\left(\mathcal{B}+\mu^{2}\right)^{2}}}
    \text{\ensuremath{\arccos}}\frac{\mathcal{B}+\mu^{2}}{2\mu}\right]\,\,\,,\label{eq:p_self_energy}\\
  n_{1}\left(\mathcal{B},\mu\right) & = &
  36-156\mathcal{B}+236\mathcal{B}^{2}-135\mathcal{B}^{3}
  +12\mathcal{B}^{4}+7\mathcal{B}^{5}+\left(72-252\mathcal{B}\right)\mu^{2}
  +\mathcal{O}\left(\mathcal{B}^{2}\right)\,\,\,,\\
  n_{2}\left(\mathcal{B},\mu\right) & = &
  -24\mathcal{B}^{3}+42\mathcal{B}^{4}-12\mathcal{B}^{5}
  +96\mathcal{B}\mu^{2}+\left(-144+348\mathcal{B}\right)\mu^{4}+\left(36-72\mathcal{B}\right)\mu^{6}
  +\mathcal{O}\left(\mathcal{B}^{2}\right)\,\,\,,\\
  n_{3}\left(\mathcal{B},\mu\right) & = &
  -24\mathcal{B}^{2}+42\mathcal{B}^{3}-12\mathcal{B}^{4}+120\mathcal{B}\mu^{2}
  +\left(-36+72\mathcal{B}\right)\mu^{4}+\mathcal{O}\left(\mathcal{B}^{2}\right)\,\,\,,
\end{eqnarray}
with $C_{N}=\left(\frac{g_{A}M_{N}}{4\pi f_{\pi}}\right)^{2}$. We
give here only terms up to $\mathcal{O}\left(\mathcal{B}^{2}\right)$
or constant in $\mu$ and list in the appendix the full coefficients
$n_{i}\left(\mathcal{B},\mu\right)$. By setting $\mathcal{B}=0$
we recover the normal nucleon $\tilde{MS}$ renormalized B$\chi$PT
self-energy \cite{ChPTDelta:partI,ChPTmasses}:
\begin{eqnarray}
  \Sigma_{p}\left(0\right) & = &
  \frac{3}{2}M_{N}C_{N}\left[1+2\mu^{2}-2\mu^{3}\sqrt{1-\frac{\mu^{2}}{4}}
    \text{\ensuremath{\arccos}}\frac{\mu}{2}-\mu^{4}\ln\mu\right]\,\,\,.\label{eq:BChPT_sigma}
\end{eqnarray}
The chiral expansion of the nucleon mass to order $p^{3}$ is 
\begin{equation}
  M_{N}\left(B\right)=\overset{\circ}{M_{N}}-4\overset{\circ}{c}_{1}m_{\pi}^{2}
  +\Sigma_{N}^{(3)}\left(B\right)-\frac{\overset{\circ}{\kappa_{N}}}{2M_{N}}B\,\,\,,
\end{equation}
where $\overset{\circ}{M_{N}}$, $\overset{\circ}{c}_{1}$ and $\overset{\circ}{\kappa_{N}}$
are the low-energy constants for the nucleon mass and its AMM. The
$\mu^{0}$ and $\mu^{2}$ terms in Eq. (\ref{eq:BChPT_sigma}) break
the usual power counting scheme \cite{BChPTLagrangian} where e.g.
the EOMS renormalization scheme \cite{EOMS} is one way to deal with
this problem. The general behavior of the self-energy in this section
is similar to the nucleon self-energy of the last section for pseudo-scalar
$N-\pi$ couplings. Especially the non-analytic $\mathcal{B}$ term
is the same and we get again an imaginary part for $\mathcal{B}<-2\mu-\mu^{2}$.

\begin{figure}
  \caption{\label{fig:proton_Sigma}Proton self-energy as function of a magnetic
    field $B$ for $m_{\pi}=139$ MeV and $M_{N}=939$ MeV. The solid
    red line is the result of Eq. (\ref{eq:p_self_energy}) and the solid
    linear green line the linear approximation $\Sigma_{p}\left(B\right)=M_{N}-\frac{\kappa}{2M_{N}}B$
    with $\kappa=1.73$. The dashed line is the imaginary part of Eq.
    (\ref{eq:p_self_energy}).}
  \includegraphics[scale=0.4]{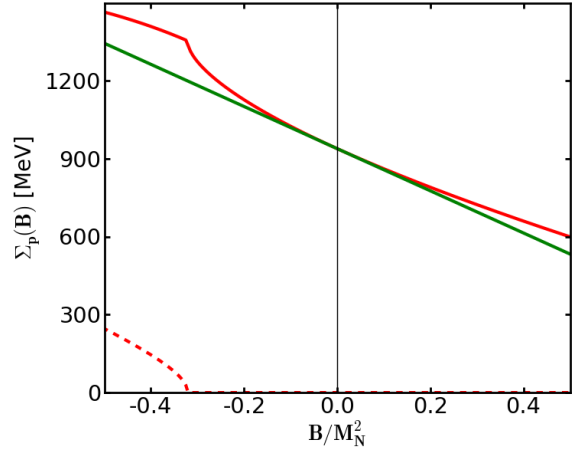}
\end{figure}
In Fig. \ref{fig:proton_Sigma} we show the proton self-energy as
function of the magnetic field $\mathcal{B}$ for the phenomenological
values. In case of the linear approximation we use the AMM obtained
from the three-point method, i.e. defined by $\kappa=F_{2}\left(0\right)$
via the matrix element

\begin{eqnarray}
  \langle N(p^{\prime})|\overline{\Psi}(0)\gamma^{\mu}\Psi(0)|N(p)\rangle & = &
  \bar{u}(p^{\prime})\left[\gamma^{\mu}F_{1}\left(q^{2}\right)+\frac{i\sigma^{\mu\nu}q_{\nu}}{2M_{N}}F_{2}\left(q^{2}\right)\right]u(p)\,\,\,,
\end{eqnarray}
with $q=p^{\prime}-p$ as the momentum transfer. The explicit results
for the nucleon graphs in the second row of Fig. \ref{fig:feynman_graphs}
are:

\begin{eqnarray}
  \kappa_{1}^{N} & = &
  \frac{1}{i}\left(\frac{g_{A}M_{N}}{2f_{\pi}}\right)^{2}\int_{0}^{1}dz2z
  \left[\left(-6+12z-4z^{2}\right)J_{2}+\left(3-4z+z^{2}\right)\varepsilon J_{2}
    -2M_{N}^{2}\left(1-z\right)^{4}J_{3}\right]\label{eq:pv_kappa1}\\
  \kappa_{2}^{N} & = &
  \frac{1}{i}\left(\frac{g_{A}M_{N}}{2f_{\pi}}\right)^{2}\int_{0}^{1}dz2
  \left[\left(3-14z+15z^{2}-4z^{3}\right)J_{2}-\left(3-7z+5z^{2}-z^{3}\right)\varepsilon
    J_{2}+2M_{N}^{2}\left(1-z\right)^{5}J_{3}\right]\label{eq:pv_kappa2}
\end{eqnarray}
with $J_{i}=J_{i}\left(\mathcal{M}\right)$ and $\mathcal{M}=zm_{\pi}^{2}+\left(1-z\right)^{2}M_{N}^{2}$
\cite{Ledwig(2011):covChptNDelta}. Extracting the AMM from the above
self-energy yields identical results where in the case of $\kappa_{1}^{N}$
this can be seen literally even before the Feynman parameter integration: 

\begin{eqnarray}
  \kappa_{1}^{N\left(BFT\right)} & = &
  \frac{1}{i}\left(\frac{g_{A}M_{N}}{2f_{\pi}}\right)^{2}\int_{0}^{1}dz2z\left[\left(-6+9z-3z^{2}\right)J_{2}
    +z\left(3-z\right)J_{2}+\left(3-4z+z^{2}\right)\varepsilon
    J_{2}-2M_{N}^{2}\left(1-z\right)^{4}J_{3}\right].
\end{eqnarray}
The two $J_{2}$ integrals add up to the same expression in Eq. (\ref{eq:pv_kappa1}).
However, in the case of the three-point function method, e.g., the
$J_{2}$ contributions come purely from tensor integrals whereas in
the case of the BFT method it is a combination of tensor and scalar
loop-integrals. As we see for the infinite volume case this difference
is not important. 

\section{Delta(1232) self-energy }

We consider now the $\Delta^{+}\left(1232\right)$-isobar and concentrate
on the graphs that give its decay width. We take the following Lagrangian
\cite{ChPTmasses}:

\begin{eqnarray}
  \mathcal{L}_{\Delta\pi} & = &
  \overline{\Delta}_{\mu}(i\gamma^{\mu\nu\alpha}D_{\alpha}-M_{\Delta}\gamma^{\mu\nu})\Delta_{\nu}
  +i\frac{h_{A}}{2f_{\pi}M_{\Delta}}\overline{N}T^{a}\gamma^{\mu\nu\lambda}\left(D_{\mu}\Delta_{\nu}\right)
  \left(D_{\lambda}^{ab}\pi^{b}\right)+\mbox{h.c.}\,\,\,,
\end{eqnarray}
and obtain the relevant self-energies as:

\begin{eqnarray}
  \Sigma_{1}^{\Delta}\left(B\right) & = &
  \frac{1}{i}M_{\Delta}\left(\frac{h_{A}}{2f_{\pi}}\right)^{2}\frac{1}{2}
  \int_{0}^{1}dz\left[\left(z+r\right)J_{1}-B\left(z+r\right)z^{2}J_{2}\right]\,\,\,,\\
  \Sigma_{2}^{\Delta}\left(B\right) & = &
  \frac{1}{i}M_{\Delta}\left(\frac{h_{A}}{2f_{\pi}}\right)^{2}\frac{1}{2}
  \int_{0}^{1}dz\left[\left(z+r\right)J_{1}-B\left(z+r\right)\left(1-z\right)^{2}J_{2}\right]\,\,\,,\\
  \mathcal{M}_{\Delta} & = &
  z\mu^{2}+\left(1-z\right)r^{2}-z\left(1-z\right)+\mathcal{B}_{i}\,\,\,,\label{eq:delta_self_energies}
\end{eqnarray}
with $\mathcal{B}_{1}=+z\left(1-z\right)\mathcal{B}$ and $\mathcal{B}_{2}=-\left(1-z\right)^{2}\mathcal{B}$
and all other definitions given in the appendix. Integrating these
expressions yield
\begin{eqnarray}
  \Sigma_{1}^{\Delta}\left(B\right)\cdot\frac{144}{M_{\Delta}C_{\Delta}} & = &
  \mathcal{A}_{1}\left(B\right)+48\left[2\left(r+\alpha_{1}\right)\lambda_{1}^{2}
    +\mathcal{B}\left(3\left(r+\alpha_{1}\right)\alpha_{1}^{2}+\left(\alpha_{1}-r\right)
    \lambda_{1}^{2}\right)\right]\,\,\Omega_{1}\left(B\right)\,\,\,,\\
  \Sigma_{2}^{\Delta}\left(B\right)\cdot\frac{144}{M_{\Delta}C_{\Delta}} & = & 
  \mathcal{A}_{2}\left(B\right)+48\left[2\left(r+\alpha_{2}\right)\lambda_{2}^{2}
    +\mathcal{B}\left(3r+3\alpha_{2}-6r\alpha_{2}-\left(3\alpha_{2}^{2}+1\right)\left(2
    +r-\alpha_{2}\right)\lambda_{2}^{2}\right)\right]\,\,\Omega_{2}\left(B\right)\,\,,\\
  \Omega_{i}\left(B\right) & = & \sqrt{\lambda_{i}^{2}}\left(\mbox{arctanh}
  \frac{\beta_{i}}{\sqrt{\lambda_{i}^{2}}}+\mbox{arctanh}
  \frac{\alpha_{i}}{\sqrt{\lambda_{i}^{2}}}\right)\,\,,\label{eq:Omega_Delta}
\end{eqnarray}
with $C_{\Delta}=\left(\frac{h_{A}M_{\Delta}}{8f_{\pi}\pi}\right)^{2}$
and the analytic parts $\mathcal{A}_{i}$ also listed in the appendix.
With these two expressions we can also write the self-energy of the
different iso-spin states as:
\begin{eqnarray}
  \Sigma_{\Delta^{++}}\left(B\right) & = & -\Sigma_{1}^{\Delta}\left(0\right)
  +\Sigma_{1}^{\Delta}\left(B\right)+\Sigma_{2}^{\Delta}\left(B\right)\,\,\,,\\
  \Sigma_{\Delta^{+}}\left(B\right) & = &
  \frac{1}{3}\Sigma_{1}^{\Delta}\left(B\right)
  +\frac{2}{3}\Sigma_{2}^{\Delta}\left(B\right)\,\,\,,\\
  \Sigma_{\Delta^{0}}\left(B\right) & = & \Sigma_{1}^{\Delta}\left(0\right)
  -\frac{1}{3}\Sigma_{1}^{\Delta}\left(B\right)+\frac{1}{3}\Sigma_{2}^{\Delta}\left(B\right)\,\,\,,\\
  \Sigma_{\Delta^{-}}\left(B\right) & = &
  2\Sigma_{1}^{\Delta}\left(0\right)-\Sigma_{1}^{\Delta}\left(B\right)\,\,\,.
\end{eqnarray}

\begin{figure}
  \caption{\label{fig:delta_sigma}The $\Delta^{+}\left(1232\right)$-isobar
    self-energy loop results as function of a magnetic field $B$ for
    $m_{\pi}=139$ MeV, $M_{N}=939$ MeV and $M_{\Delta}=1232$ MeV. The
    left pictures shows the real part and the right one the imaginary
    part. The curved red solid lines are from Eq. (\ref{eq:delta_self_energies})
    while the linear green lines correspond to the linear approximation
    $\Sigma_{\Delta}\left(B\right)=M_{\Delta}-\frac{\kappa_{\Delta}}{2M_{\Delta}}B$. }
  \includegraphics[scale=0.4]{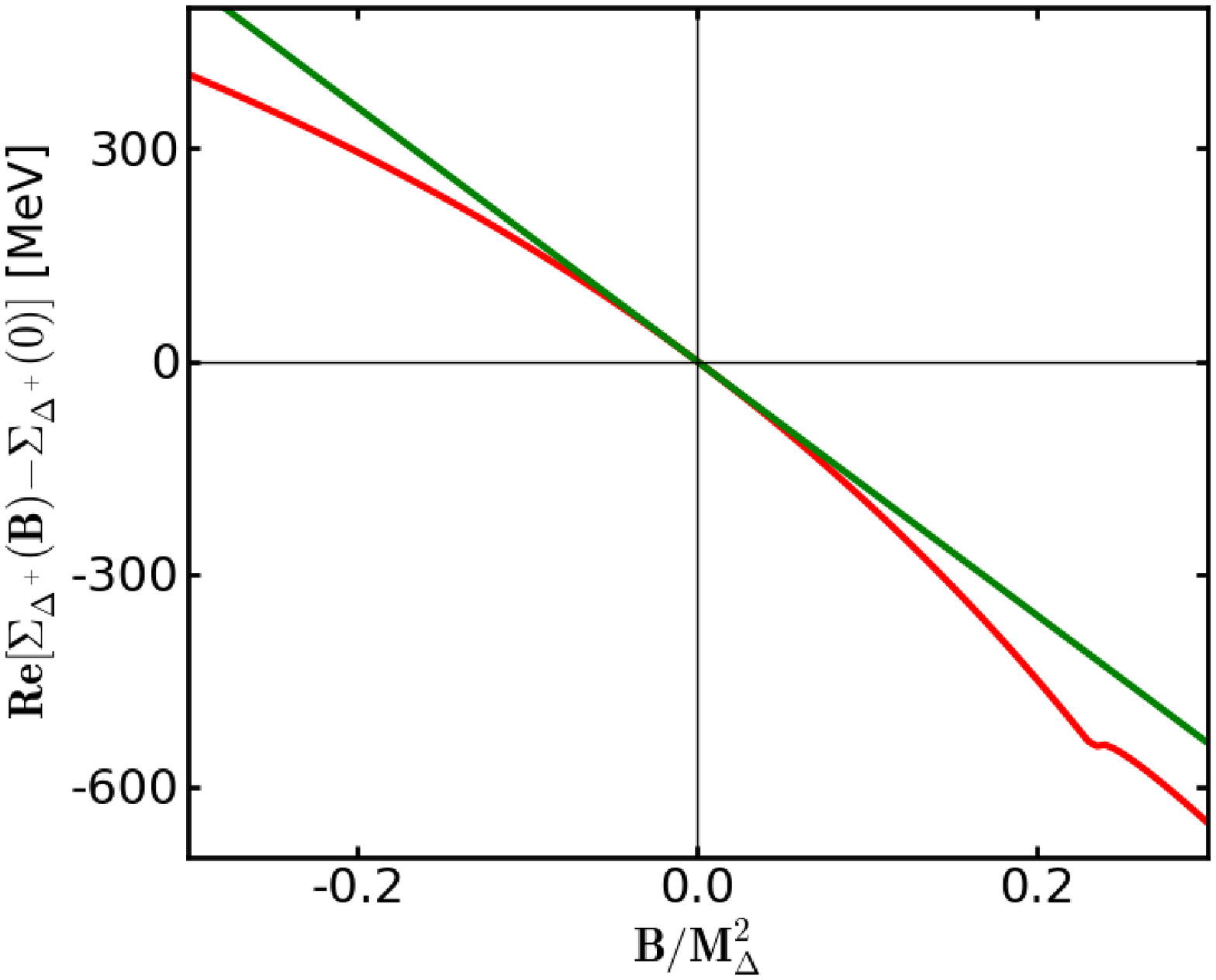}~~~~\includegraphics[scale=0.4]{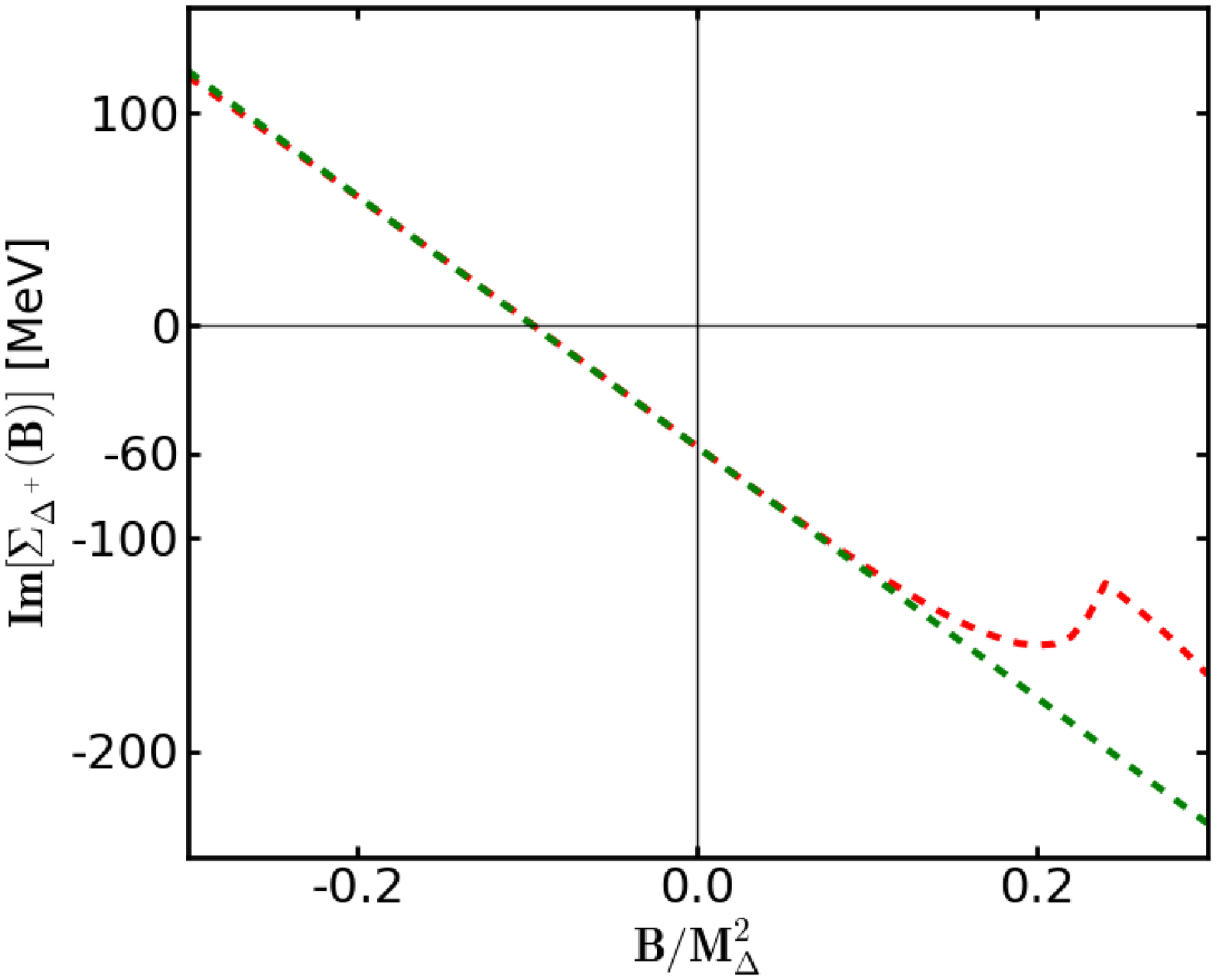}
\end{figure}
In Fig. \ref{fig:delta_sigma} we show the $\Delta^{+}\left(1232\right)$
self-energy as function of the magnetic field $B$ for the phenomenological
parameters together with the linear approximation as obtained from
the three-point function method. The cusp comes from the $\Sigma_{1}$
while the second cusp of $\Sigma_{2}$ is not present for $m_{\pi}=139$
MeV, however it emerges on the $B<0$ side for larger pion masses
and both cusps fall on $B=0$ for $m_{\pi}=M_{\Delta}-M_{N}$. The
imaginary parts read
\begin{eqnarray}
  \mbox{Im}\,\Sigma_{1}\left(B\right)\cdot\frac{3}{\pi M_{\Delta}C_{\Delta}} & =&
  -2\left(1-\mathcal{B}\right)\left(r+\alpha_{1}\right)\lambda_{1}^{3}-\mathcal{B}
  \lambda_{1}\left[3\alpha_{1}\left(\alpha_{1}^{2}+\lambda_{1}^{2}\right)
    +r\left(3\alpha_{1}^{2}+\lambda_{1}^{2}\right)\right]\,\,\,,\\
  \mbox{Im}\,\Sigma_{2}\left(B\right)\cdot\frac{3}{\pi M_{\Delta}C_{\Delta}} & =&
  -2\left(1-\mathcal{B}\right)\left(r+\alpha_{2}\right)\lambda_{2}^{3}\nn\\
  &  &-\mathcal{B}\lambda_{2}\left[-6\alpha_{2}^{2}+3\alpha_{2}^{3}-2\lambda_{2}^{2}
    +3\alpha_{2}\left(1+\lambda_{2}^{2}\right)+r\left(3-6\alpha_{2}+3\alpha_{2}^{2}+\lambda_{2}^{2}\right)\right]\,\,\,,
\end{eqnarray}
where the $\Delta^{+}\left(1232\right)$ decay width, experimentally
given by $\Gamma_{\Delta}\approx120$ MeV, is obtained from $\Gamma=-2\mbox{Im}\Sigma\left(B=0\right)$.
The slope of the imaginary part at $B=0$ in Fig. \ref{fig:delta_sigma}
is consistent with \cite{DeltaMagMom} and we obtain a vanishing
decay width at: 
\begin{eqnarray}
  \mathcal{B} & = &
  -\frac{\left(1-2r+r^{2}-\mu^{2}\right)\left(1+2r+r^{2}-\mu^{2}\right)}{2\left(1+r-2r^{3}+3r^{4}
    +2r\mu^{2}+3\mu^{4}-2r^{2}\left(1+3\mu^{2}\right)\right)}+\mathcal{O}\left(\mathcal{B}^{2}\right)\,\,\,.
\end{eqnarray}
The magnetic moment from the three-point function method is again
defined through the vector matrix element by $\kappa=F_{2}\left(0\right)$
and 

\begin{eqnarray}
  \langle\Delta(p^{\prime})|\overline{\Psi}(0)\gamma^{\mu}\Psi(0)|\Delta(p)\rangle
  & = & -\bar{u}_{\alpha}(p^{\prime})\Big\{\left[F_{1}^{\Delta}\gamma^{\mu}
    +\frac{i\sigma^{\mu\nu}q_{\nu}}{2M_{\Delta}}F_{2}^{\Delta}\right]g^{\alpha\beta}
  +\left[F_{3}^{\Delta}\gamma^{\mu}+\frac{i\sigma^{\mu\nu}q_{\nu}}{2M_{\Delta}}F_{4}^{\Delta}\right]
  \frac{q^{\alpha}q^{\beta}}{4M_{\Delta}^{2}}\Big\}u_{\beta}(p)\,\,\,,
\end{eqnarray}
with the results \cite{Ledwig(2011):covChptNDelta}:

\begin{eqnarray}
  \kappa_{1}^{\Delta} & = &
  \frac{1}{i}\left(\frac{h_{A}M_{\Delta}}{2f_{\pi}}\right)^{2}
  \int_{0}^{1}dz\,\,2z\,\,\left[-\frac{1}{2}z+z^{2}-\frac{1}{2}r+zr\right]J_{2}\,\,\,,\\
  \kappa_{2}^{\Delta} & = &
  \frac{1}{i}\left(\frac{h_{A}M_{\Delta}}{2f_{\pi}}\right)^{2}\int_{0}^{1}dz\,2
  \left(1-z\right)\left[\left(1+r\right)\left(1-z\right)-\left(1-z\right)^{2}\right]J_{2}\,\,\,,
\end{eqnarray}
with $J_{i}=J_{i}\left(\mathcal{M}_{\Delta}\right)$ and $\mathcal{M}_{\Delta}=z\mu^{2}+\left(1-z\right)r^{2}-z\left(1-z\right)$.
The results of the BFT method agree in this form literally. From $\sqrt{\lambda_{1}^{2}}$
in Eq. (\ref{eq:Omega_Delta}) we obtain the condition 
\begin{equation}
  \frac{eB}{2M_{\Delta}}\ll|M_{\Delta}-\left(M_{N}+m_{\pi}\right)|
\end{equation}
for expanding the $\Delta\left(1232\right)$-isobar self-energies
in small magnetic fields \cite{Ledwig(2010):nonAna}.

\section{Summary}

We investigated the self-energies of particles placed in a constant
electromagnetic (EM) field. Explicitly, we applied the EM background
field technique (BFT) once to the situation of stable and unstable
spin-1/2 particles coupling to (pseudo-) scalar fields and once to
the nucleon and $\Delta\left(1232\right)$-isobar baryons in the $SU\left(2\right)$
chiral perturbation theory (B$\chi$PT). We obtained the self-energies
of these particles as function of the external constant magnetic field
$B$ and calculated from these the anomalous magnetic moments (AMM)
by the linear energy shift. We summarize our findings as:
\begin{itemize}
\item Self-energies of Dirac particles coupling to (pseudo-) scalar fields:
  We investigated all three types of Feynman graphs that can appear
  in the one-loop BFT. The self-energies generally depend non-analytically
  on $B$ where for stable particles the actual non-analytic points
  are unproblematic for defining the AMM by the linear energy shift.
  These points depend on the mass and charge constellations in the loop
  and give three different conditions for resonances on when their self-energies
  can be expanded for weak $B$. One of these conditions was reported
  earlier. The self-energy formulas contain those $\mathcal{O}\left(B^{2}\right)$
  terms that keep the non-analytical dependence intact, in contrast
  to the one-photon approximation. However, they omit some $\mathcal{O}\left(B^{2}\right)$
  contributions where we estimated for the nucleon-pion system with
  pseudo-scalar couplings that these contributions are small for magnetic
  fields of $|B|<\frac{1}{5}M_{N}^{2}$ with $M_{N}$ as the nucleon
  mass.
\item Self-energies of the nucleon and $\Delta\left(1232\right)$-isobar
  in B$\chi$PT: We derived the formulas for the nucleon and $\Delta\left(1232\right)$-isobar
  self-energies depending on the pion mass and the magnetic field. We
  recover the expressions as obtained from the three point function
  method as well as the condition on when the $\Delta\left(1232\right)$-isobar
  magnetic moment is well defined by the linear energy shift. Further,
  we saw that the AMM expressions have different tensor- and scalar-loop
  integral combinations depending on whether the AMM is derived by the
  three point function method or from the self-energy. In the infinite
  volume these combinations add up to the same AMM expressions.
\end{itemize}

\begin{acknowledgments}
  The author is thankful to B. Tiburzi, V. Pascalutsa and M. Vanderhaeghen for valuable
  discussions and critical comments. 
\end{acknowledgments}

\begin{appendix}
  
\section{Notation}

We use the following notations: $\tilde{dl}=d^{n}l/\left(4\pi\right)^{n}$
with $n=4-2\epsilon$ dimensions,
\begin{eqnarray}
  \mu=m/M_{ex}\,\,\, & \,\,\, r=M_{in}/M_{ex}\,\,\, &
  \,\,\,\mathcal{B}=B/M_{ex}^{2}\,\,\,.
\end{eqnarray}
The covariant derivatives are:

\begin{eqnarray}
  D_{\mu}^{ab}\pi^{b} & = & \delta^{ab}\partial_{\mu}\pi^{b}+ieQ_{\pi}^{ab}A_{\mu}\pi^{b}\,\,\,,\\
  D_{\mu}N & = &
  \partial_{\mu}N+ieQ_{N}A_{\mu}N+\frac{i}{4f_{\pi}^{2}}\epsilon^{abc}\tau^{a}\pi^{b}
  \left(\mbox{\ensuremath{\partial}}_{\mu}\pi^{c}\right)\,\,\,,\\
  D_{\mu}\Delta_{\nu} & = &
  \partial_{\mu}\Delta_{\nu}+ieQ_{\Delta}A_{\mu}\Delta_{\nu}+\frac{i}{2f_{\pi}^{2}}\epsilon^{abc}
  \mathcal{T}^{a}\pi^{b}\left(\mbox{\ensuremath{\partial}}_{\mu}\pi^{c}\right)\,\,\,,
\end{eqnarray}
with the operators $Q_{N}=\left(1+\tau^{3}\right)/2$ and $Q_{\pi}^{ab}=-i\varepsilon^{ab3}$.
The results for the loop integration in infinite volume and dimensional
regularization with $L=-\frac{1}{\varepsilon}+\gamma_{E}+\ln\frac{M_{sc}^{2}}{4\pi\Lambda^{2}}$
is:
\begin{eqnarray}
  J_{1}\left(\mathcal{M}\right)=\frac{-i}{\left(4\pi\right)^{2}}m_{sc}^{2}
  \tilde{\mathcal{M}}\left[L-1+\ln\tilde{\mathcal{M}}\right]\,\,\,,\,\,\,&
  J_{2}\left(\mathcal{M}\right)=\frac{-i}{\left(4\pi\right)^{2}}
  \left[L+\ln\tilde{\mathcal{M}}\right]\,\,\,,\,\,\,
  &J_{3}\left(\mathcal{M}\right)=\frac{-i}{\left(4\pi\right)^{2}}
  \frac{1}{2m_{sc}^{2}}\frac{1}{\tilde{\mathcal{M}}}\,\,\,.
\end{eqnarray}
The corresponding parts of the nucleon and $\Delta\left(1232\right)$
self-energies are:

\begin{eqnarray}
  n_{1}\left(\mathcal{B},\mu\right) & = & 36-156\mathcal{B}+236\mathcal{B}^{2}
  -135\mathcal{B}^{3}+12\mathcal{B}^{4}+7\mathcal{B}^{5} \nn\\
  & & +\left(72-252\mathcal{B}+321\mathcal{B}^{2}-186\mathcal{B}^{3}
  +45\mathcal{B}^{4}\right)\mu^{2}+\left(6\mathcal{B}^{2}-6\mathcal{B}^{3}\right)\mu^{4}\\
  n_{2}\left(\mathcal{B},\mu\right) & = & -24\mathcal{B}^{3}+42\mathcal{B}^{4}
  -12\mathcal{B}^{5}+\left(96\beta-180\beta^{2}+60\beta^{3}+12\beta^{4}
  -12\beta^{5}\right)\mu^{2} \nn\\
  & & +\left(-144+348\mathcal{B}-270\mathcal{B}^{2}+120\mathcal{B}^{3}
  -18\mathcal{B}^{4}\right)\mu^{4}+\left(36-72\mathcal{B}
  +12\mathcal{B}^{2}\right)\mu^{6}+6\mathcal{B}^{2}\mu^{8}\\
  n_{3}\left(\mathcal{B},\mu\right) & = & -24\mathcal{B}^{2}+42\mathcal{B}^{3}
  -12\mathcal{B}^{4}+\left(120\mathcal{B}-294\mathcal{B}^{2}+216\mathcal{B}^{3}
-60\mathcal{B}^{4}\right)\mu^{2} \nn\\
  & &+\left(-36+72\mathcal{B}-24\mathcal{B}^{2}+6\mathcal{B}^{3}\right)\mu^{4}
-6\mathcal{B}^{2}\mu^{6}
\end{eqnarray}

\begin{eqnarray}
  \mathcal{A}\left(B\right) & = & 27+40r-68\alpha-120r\alpha+42\alpha^{2}
  +120r\alpha^{2}+12\alpha^{3}-54\lambda^{2}-168r\lambda^{2}-60\alpha\lambda^{2}\nn\\
  &  &
  +\left(-18+48\alpha-36\alpha^{2}+36\lambda^{2}-24r\left(1-3\alpha+3\alpha^{2}
  -3\lambda^{2}\right)\right)\ln\left(1-\mathcal{B}\right)\nn\\
  &  &
  +\left(-24r\alpha^{3}-6\alpha^{4}+72r\alpha\lambda^{2}+36\alpha^{2}\lambda^{2}
  +18\lambda^{4}\right)\ln\left(\alpha^{2}-\lambda^{2}\right)\nn\\
 &  &
  +\left(-18+48\alpha-36\alpha^{2}+6\alpha^{4}+36\lambda^{2}-36\alpha^{2}\lambda^{2}
  -18\lambda^{4}+24r(\alpha-1)\left((\alpha-1)^{2}-3\lambda^{2}\right)\right)
  \ln\left(\beta^{2}-\lambda^{2}\right)\\
  \mathcal{A}_{1}\left(B\right) & = & \Big[\mathcal{A}\left(B\right)+\mathcal{B}
    \Big(-36-56r+56\alpha+96r\alpha-60\alpha^{2}-168r\alpha^{2}-48\alpha^{3}
    +36\lambda^{2}+120r\lambda^{2}-48\alpha\lambda^{2}\nn\\
    &  &
    +\sqrt{\lambda^{2}}\Omega\left(144r\alpha^{2}+144\alpha^{3}-48r\lambda^{2}
    +48\alpha\lambda^{2}\right)\nn\\
    &  & +12\left(3-4\alpha+3\alpha^{2}-3\lambda^{2}+r\left(4-6\alpha+6\alpha^{2}
    -6\lambda^{2}\right)\right)\ln\left(1-\mathcal{B}\right)\nn\\
    &  & +\left(48r\alpha^{3}+24\alpha^{4}+72\alpha^{2}\lambda^{2}\right)
    \ln\left(\alpha^{2}-\lambda^{2}\right)-12\big(4\alpha+2\alpha^{4}
    +3\left(\lambda^{2}-1\right)+\alpha^{2} \left(6\lambda^{2}-3\right) \nn\\
    & &+r\left(6\left(\alpha-\alpha^{2}+\lambda^{2}\right)
    -4+4\alpha^{3}\right)\big)\ln\left(\beta^{2}-\lambda^{2}\right)\Big)
    \Big]_{\alpha,\beta,\lambda\to\alpha_{1},\beta_{1},\lambda_{1}}\\
  \mathcal{A}_{2}\left(B\right) & = & \Big[\mathcal{A}\left(B\right)+\mathcal{B}
    \Big(-40-128r+32\alpha+240r\alpha+36\alpha^{2}-168r\alpha^{2}-48\alpha^{3}
    +132\lambda^{2}+120r\lambda^{2}-48\alpha\lambda^{2}\nn\\
    &  & +12\left(2-4\alpha+3\alpha^{2}-3\lambda^{2}+r\left(4-6\alpha+6\alpha^{2}
    -6\lambda^{2}\right)\right)\ln\left(1-\mathcal{B}\right)\nn\\
    &  & +\left(72r\alpha+36\alpha^{2}-72r\alpha^{2}-48\alpha^{3}+48r\alpha^{3}
    +24\alpha^{4}+36\lambda^{2}-72r\lambda^{2}-144\alpha\lambda^{2}+72\alpha^{2}
    \lambda^{2}\right)\ln\left(\alpha^{2}-\lambda^{2}\right)\nn\\
    &  & -24(-1+\alpha)^{2}\left(-1+2r(-1+\alpha)+\alpha^{2}+3\lambda^{2}\right)
    \ln\left(\beta^{2}-\lambda^{2}\right)\Big)
    \Big]_{\alpha,\beta,\lambda\to\alpha_{2},\beta_{2},\lambda_{2}}
\end{eqnarray}

\end{appendix}

\end{document}